\documentclass[aps,
               prx,
               showpacs,
               amssymb,
               nofootinbib,
               superscriptaddress,
               twocolumn,
               longbibliography]{revtex4-2}
\usepackage{braket}
\usepackage[ansinew]{inputenc}
\usepackage{bbold}
\usepackage{bbm}
\usepackage{bm}
\usepackage{amsbsy}
\usepackage{amsthm}
\usepackage{amssymb}
\usepackage{amsfonts}
\usepackage{amsmath, mathtools}
\usepackage{dsfont}
\usepackage{graphicx}
\usepackage{epsfig}
\usepackage{epstopdf}
\usepackage{dsfont}
\usepackage{mathrsfs} 
\usepackage{multibib}
\usepackage{chemformula}	
\usepackage[colorlinks]{hyperref}
\makeatletter
\newcommand\org@hypertarget{}
\let\org@hypertarget\hypertarget
\renewcommand\hypertarget[2]{%
  \Hy@raisedlink{\org@hypertarget{#1}{}}#2%
  }
\makeatother
\usepackage[figure,table]{hypcap}

\usepackage{MnSymbol}
\usepackage{enumerate}
\usepackage{float}
\usepackage{comment}
\usepackage{braket}
\usepackage{subfigure}
\usepackage{orcidlink}
\usepackage{tcolorbox}

\newcommand{\Ss}{4^2\mathrm{S}_{1/2}}

\makeatletter
\newcommand{\providetcbcountername}[1]{%
  \@ifundefined{c@tcb@cnt@#1}{%
    --undefined--%
  }{%
    tcb@cnt@#1%
  }
}

\newcommand{\settcbcounter}[2]{%
  \@ifundefined{c@tcb@cnt@#1}{%
    \GenericError{Error}{counter name #1 is no tcb counter }{}{}%
  }{%
    \setcounter{tcb@cnt@#1}{#2}%
   }%
}%
\tcbuselibrary{theorems}
\newtcolorbox[blend into=tables]{smallboxtable}[3][]
{colback=mycolor!5, colframe=mycolor, float=ht, width=0.48\textwidth, lower separated=false, blend before title=colon hang,
title={#2}, label=#3 ,#1}
\newtcolorbox[blend into=tables]{medboxtable}[3][]
{colback=mycolor!5, colframe=mycolor, float=ht, width=0.99\textwidth, lower separated=false, blend before title=colon hang,
title={#2}, label=#3 ,#1}
\newtcbtheorem{Definitions}{Definition}%
{colback=mycolor!5,colframe=mycolor,fonttitle=\bfseries,
left=4pt,
right=4pt,
top=5pt,
bottom=5pt}{defi}
\newtcbtheorem{Theorems}{Theorem}%
{colback=mycolor!5,colframe=mycolor,fonttitle=\bfseries,
left=4pt,
right=4pt,
top=5pt,
bottom=5pt}{theorem}
\definecolor{mycolor}{rgb}{0.122, 0.435, 0.698}
\usepackage{subfigure}
\usepackage{makecell}

\hypersetup{
	bookmarksnumbered,
	pdfstartview={FitH},
	citecolor={darkgreen},
	linkcolor={darkred},
	urlcolor={darkblue},
	pdfpagemode={UseOutlines}}
\definecolor{darkgreen}{RGB}{50,190,50}
\definecolor{darkblue}{RGB}{0,0,190}
\definecolor{darkred}{RGB}{238,0,0}
\definecolor{mycolor}{rgb}{0.122, 0.435, 0.698}
\usepackage{soul}

\renewcommand{\thesubsection}{\Roman{section}.\Alph{subsection}}

\makeatletter
\renewcommand{\p@subsection}{}
\renewcommand{\p@subsubsection}{}
\makeatother

\usepackage{siunitx}
\DeclareSIUnit\inch{inch}
\DeclareSIUnit\gauss{G}

\usepackage{array}
\usepackage{colortbl}

\begin{document}

\title{Trapped-ion-based quantum repeater node across a \SI{17}{\kilo\meter} metropolitan fiber}

\author{K. Fabitha} 
\affiliation{Universit\"at Innsbruck, Institut f\"ur Experimentalphysik, Technikerstr. 25, 6020 Innsbruck, Austria}
\author{T. Runggaldier}
\affiliation{Universit\"at Innsbruck, Institut f\"ur Experimentalphysik, Technikerstr. 25, 6020 Innsbruck, Austria}
\author{A. Winkler}
\affiliation{Universit\"at Innsbruck, Institut f\"ur Experimentalphysik, Technikerstr. 25, 6020 Innsbruck, Austria}
\author{F. Bernecker}
\affiliation{Universit\"at Innsbruck, Institut f\"ur Experimentalphysik, Technikerstr. 25, 6020 Innsbruck, Austria}
\author{J. Bate}
\affiliation{Universit\"at Innsbruck, Institut f\"ur Experimentalphysik, Technikerstr. 25, 6020 Innsbruck, Austria}
\author{M.~Canteri}
\affiliation{Universit\"at Innsbruck, Institut f\"ur Experimentalphysik, Technikerstr. 25, 6020 Innsbruck, Austria}
\author{J.~Helgert}
\affiliation{Universit\"at Innsbruck, Institut f\"ur Experimentalphysik, Technikerstr. 25, 6020 Innsbruck, Austria}
\affiliation{LaserLaB, Department of Physics and Astronomy, Vrije Universiteit Amsterdam, De Boelelaan 1100, 1081 HZ Amsterdam, Netherlands}
\author{T.~E.~Northup}
\affiliation{Universit\"at Innsbruck, Institut f\"ur Experimentalphysik, Technikerstr. 25, 6020 Innsbruck, Austria}
\author{V. Krutyanskiy}
\affiliation{Universit\"at Innsbruck, Institut f\"ur Experimentalphysik, Technikerstr. 25, 6020 Innsbruck, Austria}
\author{B. P. Lanyon}
\email[Correspondence should be sent to ]{ben.lanyon@uibk.ac.at}
\affiliation{Universit\"at Innsbruck, Institut f\"ur Experimentalphysik, Technikerstr. 25, 6020 Innsbruck, Austria}
%
%

\date{\today}

\begin{abstract}

Concatenation of quantum repeater nodes has been proposed as a means to distribute entanglement efficiently across future long-distance networks. 
Here, we demonstrate a trapped-ion repeater node that establishes photonic entanglement over a \SI{17}{\kilo\meter}  fiber traversing a city,  moving beyond laboratory-only prototypes. The node comprises two atomic ions coupled to an optical cavity. First, a detailed characterization is performed of the long fiber channel and of ion-photon entanglement distributed across it. Next, a repeater-node protocol is implemented, achieving entanglement between a local photon and a remote photon with a Bell-state fidelity of $81.6^{+1.9}_{-1.4}\%$. These results demonstrate that trapped-ion repeater-node technology can operate over existing city-scale fiber infrastructure and represent a step toward distributed quantum repeater networks.\\

\end{abstract}

\date{\today}

\maketitle


\addtocontents{toc}{\protect\iffalse}

Significant efforts are underway to develop light-matter quantum networks, consisting of distributed nodes---each housing matter-based quantum registers for storing and processing quantum information---interconnected by photonic links \cite{Kimble2008, WehnerElkoussHanson2018}. Such entanglement-based networks, operating over distances ranging from citywide to continental scales, would enable powerful applications in secure communication \cite{PirandolaEtAl2020}, distributed computing \cite{JiangTaylorSorensenLukin2007}, precision sensing \cite{Zhang_2021}, and timekeeping \cite{Komar2014}. Prototype city-scale light-matter networks are already under development \cite{RakonjacGrandiWengerowsky2023, StolkvanderEndenSlater2024, BersinGreinSutula2024, kucera2024demonstration, LiuLuoYu2024, nc84-zjp3}. A central strategy for extending the reach of these emerging networks is the implementation of quantum repeaters: 
devices that enable long-distance quantum communication via entanglement swapping and purification \cite{DuerBriegelCiracZoller1999}. 
Various architectures for quantum repeaters have been developed, and extensive experimental progress toward their realisation has been made across diverse physical platforms \cite{Azuma2023}. 

\begin{figure}[t]
	\centering
	\includegraphics[width=0.9\columnwidth]{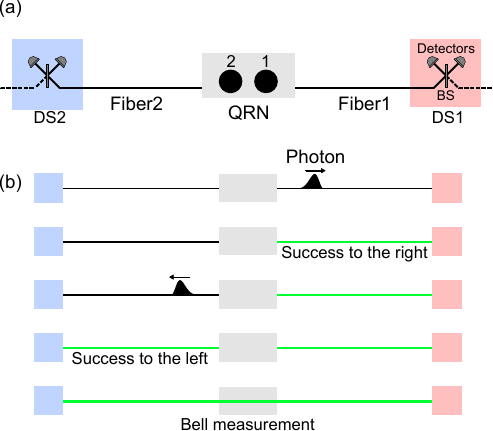} 
	\caption{
	\textbf{Quantum repeater node (QRN) concept.} 
(a) The QRN contains at least two matter-based qubits (e.g., atoms) and lies between two optical fibers. QRN-emitted photons are detected at the two detection stations (DS), each of which contain a beam splitter (BS) and photon detectors. Dashed lines show extensions to concatenate QRNs to build repeater chains: photon detection heralds entanglement between atoms in remote QRNs. (b) QRN protocol: 1) Attempts to distribute a photon entangled with atom 1 to DS1 are repeated until success (green line). 2) While atom 1 stores a qubit in memory, attempts to distribute a photon entangled with atom 2 to DS2 are repeated until success (green line). 3) A Bell-state measurement on the atomic qubits swaps entanglement onto the two asynchronously distributed photons.}\label{figure1}
\end{figure}

Trapped ions could serve as a powerful platform with which to realize quantum repeaters \cite{Sangouard2009}, owing to record coherence times and high-fidelity gates that enable deterministic entanglement swapping, among other advantages.
Recently, individual quantum repeater nodes (QRNs) based on trapped ions have been demonstrated~\cite{KrutyanskiyCanteriMeraner2023,PhysRevA.110.032603}. 
A QRN is a device inserted into a fiber channel that contains at least two photon-interfaced matter-based (e.g., atomic) qubits~\cite{Luong:2016dd, Weinfurter2020_theory_repeater, Langenfeld_Rempe_21}. By combining on-demand generation of atom-photon entanglement, storage in atomic memories, and a Bell-state measurement on the atoms, a QRN establishes photonic entanglement between the ends of the fiber channel (Fig.~\ref{figure1}). Multiple QRNs could in the future be concatenated into a repeater chain to extend entanglement over longer distances; in such concatenated chains, photon-detection events serve as heralds that signal successful establishment of long-distance matter-matter entanglement~\cite{Weinfurter2020_theory_repeater,KrutyanskiyCanteriMeraner2023}. To date, trapped-ion-based QRN demonstrations have been performed in laboratory settings using spooled optical fibers. Demonstrating comparable performance in deployed scenarios, with photons distributed outside controlled laboratory environments, has so far remained an outstanding challenge.

In this work, we demonstrate the trapped-ion quantum repeater node of \cite{KrutyanskiyCanteriMeraner2023} now with one of its fiber segments deployed across and outside of a city. 
Specifically, the repeater node establishes entanglement between a photon detected after traveling through a deployed \SI{17}{\kilo\meter} fiber and a second photon, detected in the repeater node's lab after traveling through a fiber a few meters long. The rest of this paper is structured as follows. First, the experimental setup is described. Second, the deployed fiber is characterized in terms of its transmission efficiency, polarization dynamics and  light pollution introduced. Third, the fidelity, efficiency and rate with which ion-photon entanglement is distributed across the deployed fiber are investigated. Fourth, the repeater-node results are presented, corresponding to the establishment of distributed entangled photon-photon states.  
Finally, there is a short discussion and outlook.

\begin{figure}[t]
	\centering
	\includegraphics[width=1\columnwidth]{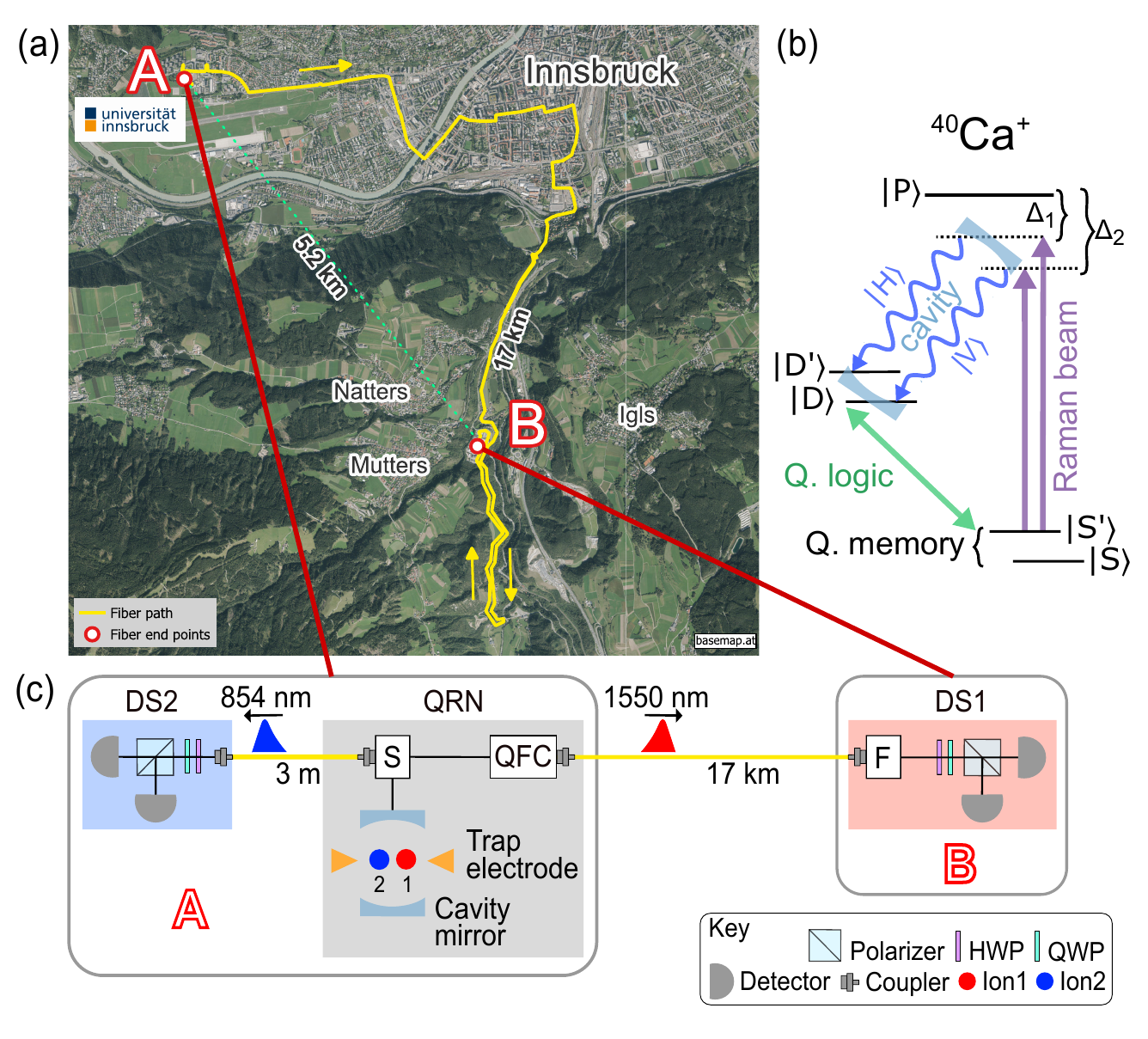}
	\caption{
	\textbf{Quantum network}. (a) Satellite image generated using Ref.~\cite{basemap_at,QGIS}. Locations A and B are two buildings separated by \SI{5.2}{\kilo\meter} (dashed green line) and connected by two \SI{17}{\kilo\meter}-long optical fibers (yellow line). (b) Energy-level diagram of $^{40}\mathrm{Ca}^{+}$ showing photon generation process. The relevant Zeeman states are $\ket{S}{=}\ket{4^{2} S_{1/2,m_j{=}{-}1/2}}$, $\ket{S'}{=}\ket{4^{2} S_{1/2,m_j{=}+1/2}}$, $\ket{P}{=}\ket{4^{2} P_{3/2,m_j{=}-3/2}}$, $\ket{D}{=}\ket{3^{2}D_{5/2,m_j{=}-5/2}}$, and $\ket{D'}{=}\ket{3^{2}D_{5/2,m_j{=}-3/2}}$. Bichromatic Raman drive, detuned by $\Delta_1$ and $\Delta_2$ from the excited state, generates  $\ket{V}$ and $\ket{H}$ polarized photons. Ion-qubit quantum logic operations are done on the $\ket{S}$--$\ket{D}$ transition (green arrow). (c) Experimental setup. In QRN at Location A, $^{40}\mathrm{Ca}^{+}$ in a cavity-integrated linear Paul trap generates photons at \SI{854}{\nano\meter}. A switch (S) directs photons either to a quantum frequency conversion (QFC) stage (\SI{854}{\nano\meter} to \SI{1550}{\nano\meter}) or to DS2 for polarization analysis (waveplates, polarizer, and fiber-coupled detectors). At Location B, the \SI{1550}{\nano\meter} photons are received in DS1, where the photons pass through a fiber Bragg grating (F) before polarization analysis.
	}
		\label{fig1_map_and_intro}
\end{figure}

\section{Experimental setup}

Our network consists of a \SI{17}{\kilo\meter} optical fiber path that connects a laboratory in the city of Innsbruck (Location A) to an office located in a suburb (Location B), as shown in Fig.~\ref{fig1_map_and_intro}(a). We rent two fibers along this path from the local municipal services company. One is used to transport  single photons from Location A to B and is henceforth referred to as Channel Q. The other is used in the other direction to return classical (laser) signals and is referred to as Channel C. Both fibers are single mode (SMF-28) and suitable for transmission in the telecom C-band, centered around \SI{1550}{\nano\meter}. The fibers are part of the preexisting fiber infrastructure in Innsbruck and are dark in the sense that they are used for our purposes exclusively. A detailed description of the experimental setup is presented in Appendix \ref{appendix:experimental setup}.

The lab setup (Fig.~\ref{fig1_map_and_intro}(c)) at Location A includes a linear Paul trap with an integrated \SI{20}{\milli\meter} optical cavity for the collection of ion-emitted single photons at \SI{854}{\nano\meter} \cite{Canteri2025}. Up to two $^{40}$Ca$^{+}$ ions are co-trapped in this work, each positioned near the maximum of a standing wave of the optical cavity at \SI{854}{\nano\meter}.  
A photon can be generated from either one of the atoms via a bichromatic cavity-mediated Raman transition (Fig.~\ref{fig1_map_and_intro}(b))~\cite{Stute2012}. Here, a 393 nm laser beam, focused to an intensity waist of \SI{1.2}{\micro\meter} on the selected ion, drives the transition and ideally prepares the maximally entangled ion-photon Bell state
$\ket{\psi(\phi)}=1/\sqrt{2}\left (|\text{D}\text{V}\rangle  + e^{-i\phi} |\text{D}'\text{H}\rangle\right)$, where $|\text{D}\rangle$ and $|\text{D}'\rangle$ are the respective metastable Zeeman states $|3^2\text{D}_{5/2}, m_j = -5/2\rangle$ and $|3^2\text{D}_{5/2}, m_j = -3/2\rangle$, $|\text{V}\rangle$ and $|\text{H}\rangle$ are the vertical and horizontal polarization components of a photon emitted into the cavity vacuum mode, and $\phi$ is a tunable phase \cite{Stute2012}. After the photon exits the cavity through the output mirror, it enters a single-mode fiber-coupled optical switch and is sent either to a polarization analysis setup (DS2, Fig.~\ref{fig1_map_and_intro}(c))  or to a quantum frequency conversion (QFC) setup, which are both described below. An acousto-optic deflector shifts the focal point of the Raman laser to the co-trapped ion within a few microseconds, allowing for a second ion-entangled photon to subsequently be generated. 

A focused laser beam at \SI{729}{\nano\meter}, which ideally couples equally to both ions, is used both to transfer ion qubits between different encodings and to perform quantum logic operations, as now described. For ion-qubit storage on the millisecond timescale, the metastable qubit ($\ket{D}$/$\ket{D'}$) is transferred to the ground state qubit ($\ket{S}$/$\ket{S'})$, which serves as a quantum memory, see Fig.~\ref{fig1_map_and_intro}(b). Dynamical decoupling extends the coherence time of the ground state qubit, realized with radio frequency (RF) pulses at \SI{11.6}{\mega\hertz} generated by a current-carrying coil that lies outside the vacuum chamber. Entangling quantum logic gates are operated on optical qubits, encoded in superpositions of $\ket{S}$ and $\ket{D}$. 
Specifically, a \SI{729}{\nano\meter} laser pulse of \SI{205}{\micro\second} length realizes a two-qubit entangling M\o lmer--S\o rensen (MS) gate \cite{PhysRevLett.82.1971, PhysRevA.62.022311, Benhelm:2008xn} via off-resonant driving of the outermost radial center-of-mass (COM) motional mode of the ion string. 
Ion-qubit readout is performed on optical qubits and consists of optional \SI{729}{\nano\meter} single-qubit rotations to set the desired measurement basis, followed by state detection via
electron shelving \cite{roos2004bell}. 
The fluorescence state of each ion (bright or dark) is determined individually, via imaging onto a digital camera, taking \SI{2.5}{\milli\second}.

\begin{figure}[t]
	\centering
	\includegraphics[width=\columnwidth]{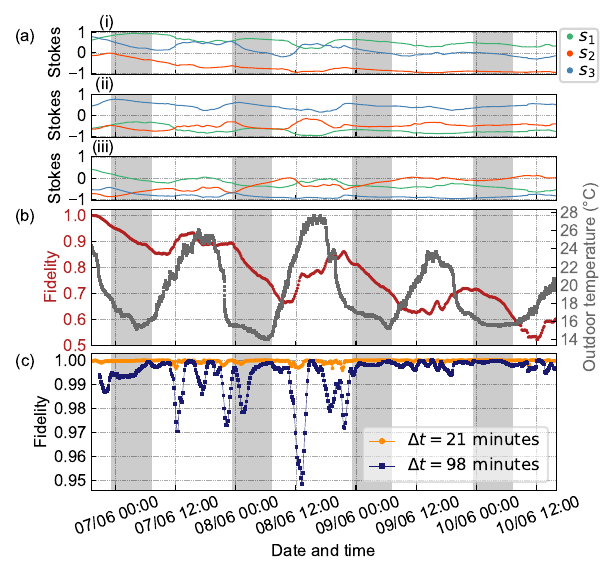} 
	\caption{
	\textbf{Polarization dynamics of laser light through Channel C.} 
	Gray shaded regions indicate nighttime. 
	(a)~Output Stokes parameters, for input polarizations: (i)~horizontal, (ii) diagonal and (iii) right-circular. 
	(b) Process fidelity $F(E(t), E(t_0))$ between the polarization process implemented by Channel C at time $t$ and the first process measured at time $t_0$ (red line, left axis). Gray data points (right axis) show the outdoor temperature measured in downtown Innsbruck \cite{ACINN2025}. 
	(c) Process fidelity $F(E(t), E(t-\Delta t))$ between measurements acquired 21 minutes apart (yellow circle) and 98 minutes apart (blue square).  	}
	\label{figure2}
\end{figure}

The polarization analysis setup at Location A (DS2, Fig.~\ref{fig1_map_and_intro}(a)) consists of a half-wave plate (HWP), a quarter-wave plate (QWP), a polarizing beam splitter and two fiber-coupled superconducting nanowire single-photon detectors (SNSPDs) optimized for \SI{854}{\nano\meter}. The QFC setup converts incoming fiber-coupled \SI{854}{\nano\meter} photons to outgoing fiber-coupled \SI{1550}{\nano\meter} photons, in a polarization-preserving fashion. The conversion process is realized via difference frequency generation involving a \SI{1901}{\nano\meter} pump laser and a periodically poled LiNbO$_3$ crystal in Sagnac configuration~\cite{Bock2018}. More details on the QFC setup can be found in Appendix \ref{appendix:QFC Setups and Characterization}. The converted telecom-wavelength photons are then injected into Channel Q with a fiber joiner. 

The setup at Location B (Fig.~\ref{fig1_map_and_intro}(c)) includes a similar photon polarization analysis setup to that at Location A but optimized for \SI{1550}{\nano\meter} photons (DS1). At the input is an optical bandpass filter (fiber Bragg grating, FBG) centered at \SI{1550}{\nano\meter}, with a specified bandwidth of \SI{1.7(3)}{\giga\hertz}. Detector clicks are converted to optical signals via an electro-optic coupler and sent back to Location A via Channel C. The received signals at Location A are recorded on a timetagger and used to trigger readout of the ion-qubit states.  
Single photons, or return optical signals, require \SI{82.0}{\micro\second} to traverse the fiber paths between Locations A and B. This travel time is calculated using a refractive index  of 1.468 at 1550 nm for the 16.74 km-long  SMF-28 fiber of Channel Q. 

Appendix ~\ref{appendix:setup-polstab} provides additional details on our setup for characterizing and stabilizing the polarization dynamics in the channels; this setup is not shown in Fig.~\ref{fig1_map_and_intro}(c). In summary, a flip mirror at the input of a channel allows polarized \SI{1550}{\nano\meter} laser light to be injected at Location A, while a second flip mirror at the output directs the light to a polarimeter at Location B. Desired input polarization states are prepared using waveplates, and both the waveplates and flip mirrors are remotely controlled.

\section{Channel characterization}\label{Channel_characterization}

This subsection presents experiments characterizing relevant properties of the deployed fibers prior to single-photon transmission. Transmission efficiency and polarization dynamics of laser light sent through the fibers are reported, followed by quantification of in-coupled background light at the single-photon level.

\noindent\textbf{Transmission efficiency.}
The transmission efficiency of Channel Q is determined by injecting \SI{1550}{\nano\meter} laser light into its input in the lab and measuring the power directly at its output at Location B.
The result is a ratio of output to input power of $P_5=0.192(3)$, which is taken as the transmission efficiency for Channel Q in all subsequent calculations. 
Separate measurements of individual sections of Channel Q (Appendix ~\ref{appendix:setup-Channel}) were performed by external companies using optical time-domain reflectometry (OTDR). Summing the reported section losses yields a higher transmission efficiency of 0.239 for Channel Q.
The OTDR-derived value does not include losses due to connecting the individual sections together and due to coupling light into the first section. 
A typical attenuation rate for \SI{1550}{\nano\meter} light traversing SMF-28 fiber is \SI{0.2}{\decibel/\kilo\meter}, which would yield a transmission efficiency over the \SI{16.74}{\kilo\meter} length of Channel Q of $0.463$.
The lower measured efficiency ($P_5$) is attributed to losses in the splices and connectors along Channel Q.  
A typical attenuation rate for the ion-emitted wavelength of \SI{854}{\nano\meter} in wavelength-optimized standard silica fiber is \SI{3}{\decibel/\kilo\meter}, which would yield a transmission efficiency over the Channel Q length of $1\times 10 ^{-5}$. The QFC setup thus plays a critical role in converting the photon wavelength to \SI{1550}{\nano\meter}.

\begin{figure}[t]
	\centering
	\includegraphics[scale=0.3]{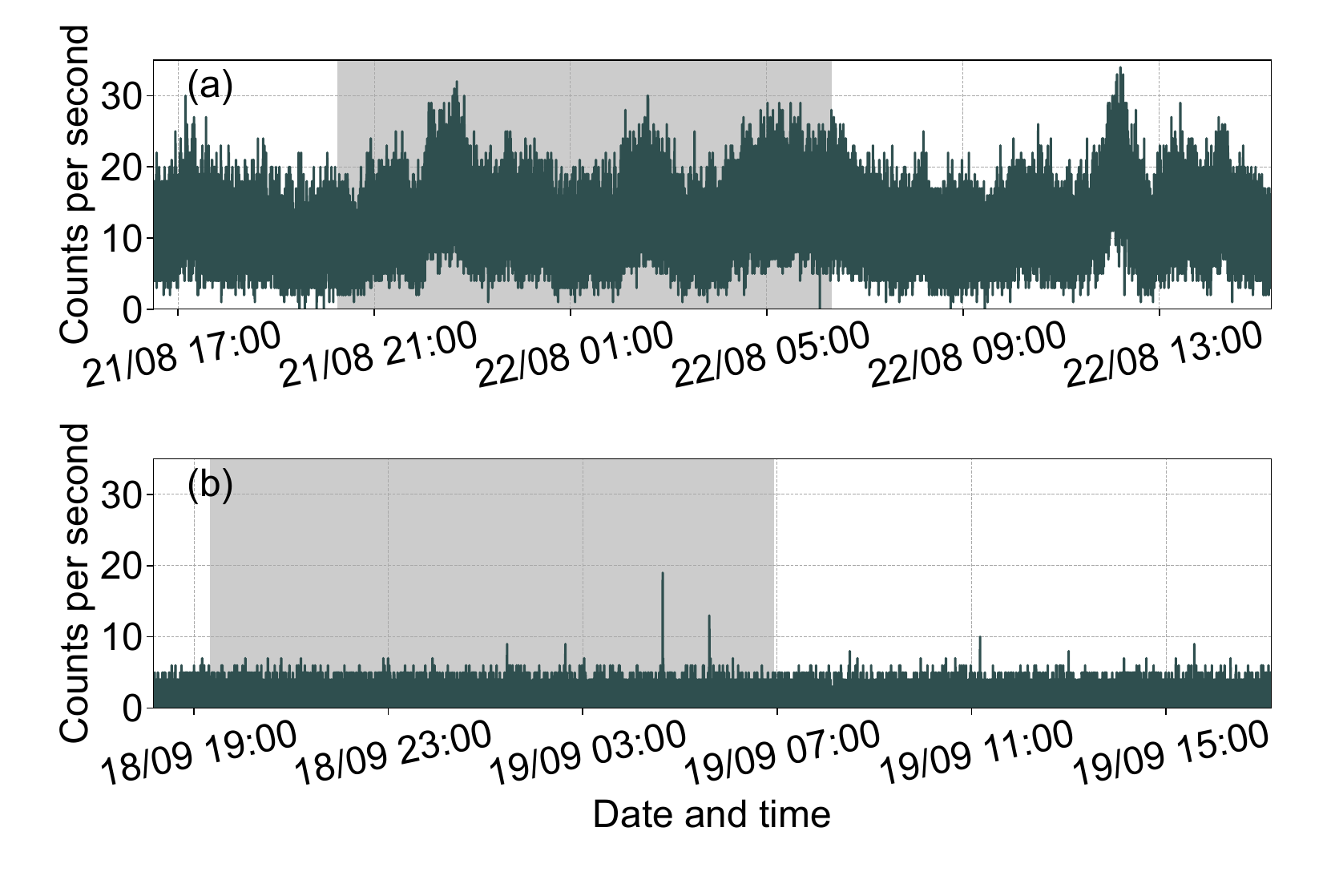}
	\caption{
		\textbf{Background photons introduced by Channel Q.}
		Total single-photon detector counts from both detectors at Location B when no light is injected into Channel Q at Location A. (a) Without the fiber Bragg grating (FBG) filter: filtering bandwidth \SI{12}{\nano\meter} (\SI{1.5}{\tera\hertz}). (b) With the FBG in place: filtering bandwidth \SI{14(2)}{\pico\meter} (\SI{1.7(3)}{\giga\hertz}). Nighttime is shaded in gray.}
	\label{figure3}
\end{figure}

\noindent\textbf{Polarization stability.}
Since we encode qubits in photon polarization, time-varying polarization rotations introduced by the optical fibers during a measurement run can degrade the measured state fidelity. In optical fibers, polarization rotations can change over time due to dynamically evolving stresses and strains modifying the fiber's birefringence. 
We study the dynamical evolution of the polarization of laser light transmitted through one of the deployed fibers (channels) over a period of several days. During this time, we repeatedly performed tomography of the process \cite{NielsenChuang2010} implemented by the respective channel on the polarization of transmitted \SI{1550}{\nano\meter} laser light, as summarized below. More details on the experimental setup and the methods for characterizing and stabilizing the polarization in deployed fibers are given in Appendix ~\ref{appendix:setup-polstab}.

Each process tomography run consisted of injecting \SI{1550}{\nano\meter} laser light with six different polarizations into the channel and, in each case, measuring the output Stokes parameters using a polarimeter. The corresponding Choi process matrix $E$ \cite{HradilEtAl2004} was obtained using the maximum likelihood technique. Obtaining each process matrix $E$ took 28 seconds, almost all of which was spent rotating the waveplates to prepare the different input states. 
The similarity of two process matrices measured at different times, $E_1$ and $E_2$, is quantified by the process fidelity $F(E_1,E_2)=\left(\mathrm{tr}\left(\sqrt{\sqrt{E_1}\,E_{2}\,\sqrt{E_1}}\right)\right)^2$~\cite{PhysRevA.71.062310}.

Results are now presented from process tomography runs performed at ten-minute intervals over a period of four days through Channel C. For the subsequent experiments involving single photons, Channel Q was used owing to its slightly higher transmission efficiency, as indicated by OTDR measurements.
The recorded Stokes parameters at the output of Channel C for three example input polarizations are presented in Fig.~\ref{figure2}(a). 
All reconstructed process matrices are well described by unitary rotations. Specifically, the fidelity $F$ between each process matrix and its nearest single-qubit unitary operator $U$ lies between 0.994 and 0.998. We now consider the \emph{initial process fidelity} $F(E(t),E(t_0))$, where $E(t)$ is the process matrix recorded at time $t$ and $E(t_0)$ is the first one measured. Figure~\ref{figure2}(b) compares the dynamics of this initial process fidelity with the outdoor air temperature measured near Location A. The initial process fidelity oscillates with a period of approximately one day. A partial recovery of the fidelity starts when the temperature returns to a value similar to the temperature at the initial time $t_0$. 

To better assess the dynamics over shorter timescales, we next consider the fidelity between processes separated by a time $\Delta t$: $F(E(t),E(t-\Delta t))$. When considering processes 21 minutes apart, the fidelity remains greater than 0.99 in $100\%$ of the cases over four days (Fig.~\ref{figure2}(c)).
Even for processes measured 98 minutes apart, a fidelity of greater than 0.99 is observed in $83\%$ of the cases.

The results presented here constitute a thorough characterization of the polarization dynamics in Channel C. A shorter polarization stability measurement on Channel Q over almost 2.5 days showed that the fidelity $F(E(t),E(t-\Delta t))$ between processes separated by $\Delta t = 21$\,minutes exceeds 0.99 in $100\%$ of the cases and for $\Delta t = 98$\,minutes exceeds 0.99 in $89\%$ of the cases; it is therefore comparable to the fidelity of Channel C reported above.

Polarization rotations imparted by an optical fiber can be stabilized by first characterizing them and then correcting them via feedback to controlled birefringent elements, as demonstrated in, e.g., Refs.~\cite{kucera2024demonstration, treiber2009fully, neumann2022continuous,strobel2024,PRXQuantum.5.030330}. 
Given the stability of our deployed fibers, we chose to stabilize polarization at average intervals of 20 minutes and 1.5 hours in the experiments presented in Sec.~\ref{ion-photon entanglement} and Sec.~\ref{Repeater node demonstration}, respectively. Each stabilization run involved two steps. In the first step, we characterized the process, $E(t)$, implemented by Channel Q on the polarization of injected \SI{1550}{\nano\meter} laser light at time $t$. We then calculated the nearest unitary rotation $U(t)$ to $E(t)$. In the second step, we set the waveplates immediately before the input of Channel Q (Fig.~\ref{fig1_map_and_intro}(c)) to implement the inverse $U^{-1}(t)$, such that the combination of waveplates and Channel Q ideally realized the operator $U^{-1}(t)U(t)=\mathbb{1}$, where $\mathbb{1}$ is the one-qubit identity operator. The characterization step takes \SI{28}{\second} seconds, and the correction step took up to \SI{8}{\second}, yielding a total time for each stabilization run of up to \SI{36}{\second}. This time was entirely dominated by waveplate rotation time and could be significantly sped up if desired using birefringent elements with faster control (e.g., see \cite{kucera2024demonstration}). In between stabilization runs, experiments involving single photon transmission were carried out. 
 
\begin{figure}[t]
	\centering
	\includegraphics[scale=0.13]{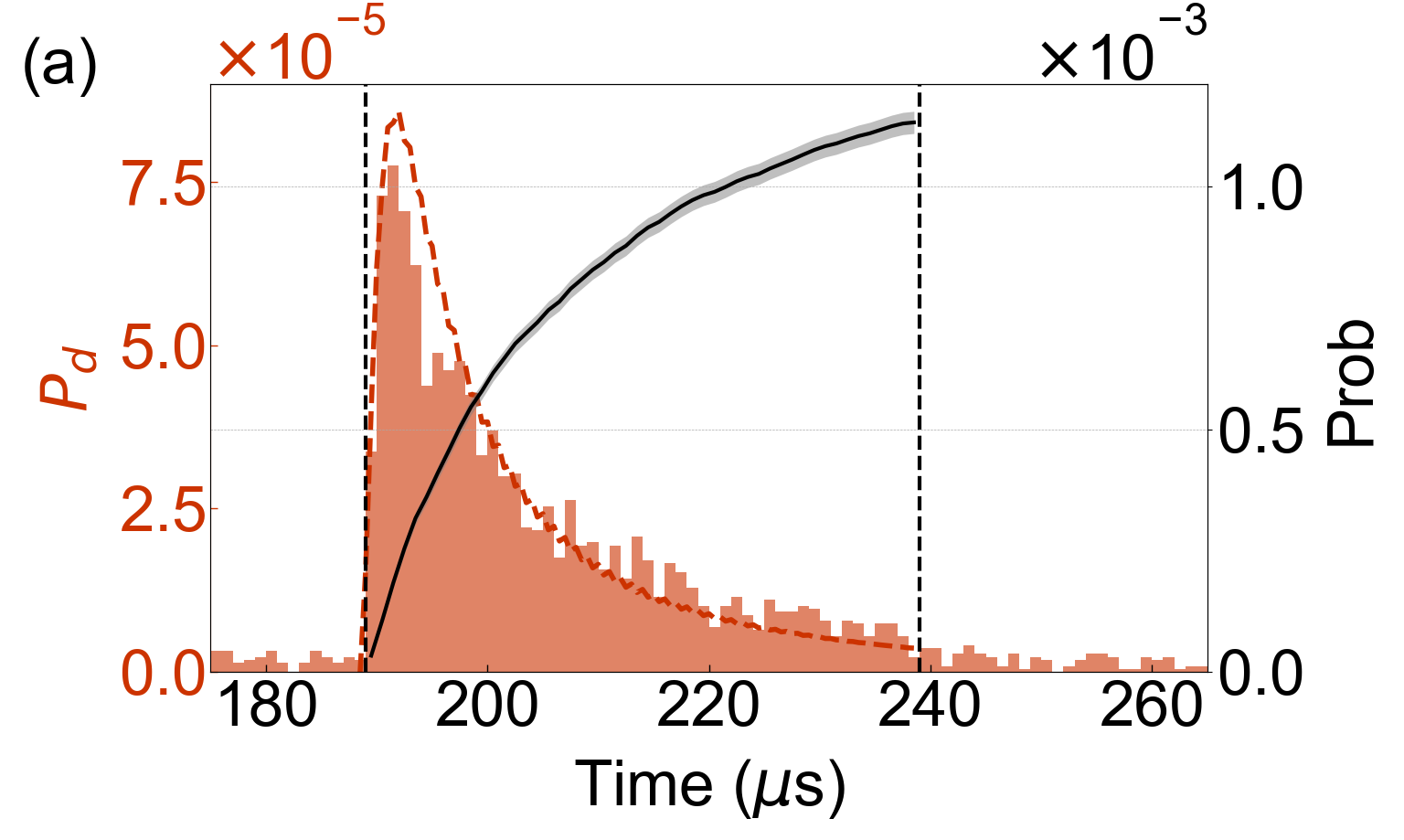}
	\includegraphics[scale=0.10]{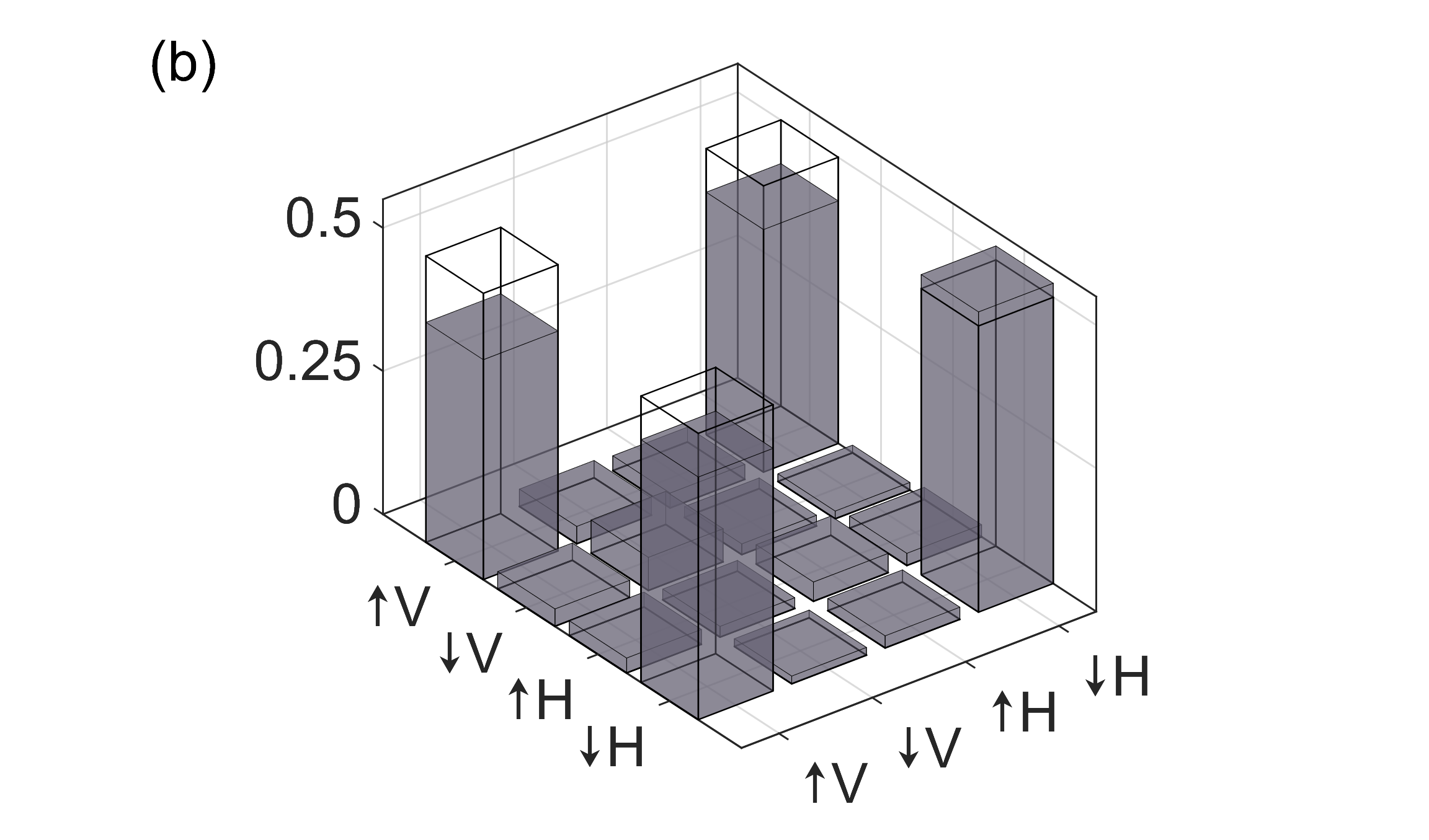}
	\caption{
	\textbf{Ion-photon entanglement across the network}. 
	(a) Histogram of single photon detection events at Location B. Probability density is shown on the left vertical axis: the number of counts normalized by the number of attempts and by the time-bin width of \SI{1}{\micro\second}. Red bars show data. Dashed red line shows a theoretical model and the solid black line (right vertical axis) represents cumulative photon detection probability with statistical uncertainty shown by gray shading. Quantum state tomography of the joint ion-photon state is performed considering detection events within the window indicated by the dashed black lines.  
	(b) Absolute value of the density matrix of the reconstructed joint two-qubit ion-photon state. The wireframe shows the ideal Bell state.}
	\label{fig:ion-photon}
\end{figure}

\noindent\textbf{Background light.} 
Light can couple into an optical fiber from sources along its path, such as room lights and data traffic in neighboring fibers. This light pollution can lead to photons exiting the fiber that lie within the final filtering bandwidth, acting as a background to the signal photons emitted by the ion. We measured this background using the setup in Fig.~\ref{fig1_map_and_intro}(c), by blocking the input of Channel Q in the lab and recording the photon detector clicks of both detectors at Location B. 

In the first measurement run, the background counts were recorded over a day without the fiber Bragg grating (FBG) filter at the output of Channel Q. The filtering bandwidth in this case was set by the remaining \SI{12}{\nano\meter} bandpass filters centered at \SI{1550}{\nano\meter}. The recorded counts summed over both detectors, shown in Fig.~\ref{figure3}(a), have a mean value of 10.80(1) cps.  In the second measurement run, the FBG filter was included which has a bandwidth of \SI{14(2)}{\pico\meter} at \SI{1550}{\nano\meter}. The recorded counts summed over both detectors, shown in Fig.~\ref{figure3}(b), have a reduced mean value of 1.208(4) cps which is consistent with the sum of intrinsic detector dark counts. 
Therefore, within the filtering bandwidth, we do not detect any background photons introduced by Channel Q. This filtering bandwidth is four orders of magnitude broader than the bandwidth of photons that exit our cavity and therefore should not introduce spectral-filtering-induced photon loss.

\section{Distributed ion-photon entanglement}\label{ion-photon entanglement}

\noindent\textbf{Experimental details.} 
The experimental sequence begins with state initialization of a single trapped ion by Doppler cooling and optical pumping, which takes \SI{3.24}{\milli\second}.  
The sequence continues with a loop in which up to 10 attempts are made to establish remote ion-photon entanglement. In each iteration of the loop, optical pumping (\SI{30}{\micro\second}) and metastable-state repumping pulses (\SI{60}{\micro\second}) are followed by a Raman laser pulse of length \SI{50}{\micro\second} which ideally generates the ion-photon entangled state $\ket{\psi(\phi)}$. The photon is ideally converted to \SI{1550}{\nano\meter} in the QFC setup, sent into Channel Q and detected at Location B. In the case in which either of the two photon detectors at Location B fires within the \SI{50} {\micro\second} arrival time window, a signal is sent back to Location A via Channel C (\SI{82.0} {\micro\second} travel time), the sequence exits the loop, and the ion qubit is measured. 
Ion-qubit measurement begins by mapping the metastable qubit $\ket{D}$/$\ket{D'}$  to the optical qubit $\ket{S}$/$\ket{D'}$  (\SI{5.2}{\micro\second}), followed by an optional laser pulse to change measurement basis (\SI{3.1}{\micro\second}) and state detection via electron shelving (\SI{2.5} {\milli\second}). 
After the loop of up to 10 attempts ends, the sequence restarts again, beginning with state initialization. 

The remote ion-photon state is characterized via quantum state tomography, for which the experimental sequence is repeated for different measurement bases of ion and photon. The measurement basis for the detected photon polarization qubit is set by the angles of the waveplates at Location B. 
The measurement basis for the ion qubit is set by \SI{729}{\nano\meter} single-qubit rotations. 
For state tomography, we measure in 18 ion-photon bases, corresponding to all combinations of the three Pauli bases for the ion and the six polarization bases $[H/V, V/H, +/-, -/+, R/L, L/R]$, where $+/-$ are diagonal/antidiagonal basis and $R/L$ are right-/left-circular basis. Swapped bases, e.g., $H/V$ and $V/H$, are measured by swapping, e.g., the detectors for the H and V photons via choice of waveplate setting, which allows for cancellation of any detector efficiency imbalances in the subsequent analysis. 
Tomographic reconstruction, via the maximum likelihood technique \cite{PhysRevA.64.052312}, yields the two-qubit ion-photon density matrix $\sigma^{\text{B}}_{\text{ip}}$, where the B superscript specifies that the photon is detected at Location B. 
Statistical uncertainties in properties derived from the density matrix are estimated via Monte Carlo resampling \cite{efron1986_MC}, based on Poissonian photon counting statistics. 

We quantify the entanglement in a two-qubit state via the concurrence $C$ \cite{HillWootters1997}. 
The density matrix $\sigma^{\text{B}}_{\text{ip}}$ is expected to deviate from the ideal state $\ket{\psi(\phi)}$, e.g., due to unitary rotations of the photon polarization imparted by the fibers between the vacuum chamber and the input of Channel Q. To account for this, we numerically search for the single-qubit rotations which maximize the fidelity with the Bell state $\ket{\psi(0)}$. That is, we maximize $\bra{\psi(0)}U\sigma^{\text{B}}_{\text{ip}}U^{\dagger}\ket{\psi(0)}$, where $U=u_\text{i}\otimes u_\text{p}$ is a tensor product of single-qubit rotations on the ion and photon, respectively.  For the value of $U$ that maximizes this expression, we then calculate $\rho^{\text{B}}_{\text{ip}}=U\sigma^{\text{B}}_{\text{ip}}U^{\dagger}$. 
These rotations do not change the entanglement content of the measured state and allow for the state to be easily visualized and for calculation of the fidelity. 
The fidelity of our state with respect to the nearest maximally entangled state is $F=\bra{\psi(0)}\rho^{\text{B}}_{\text{ip}}\ket{\psi(0)}$. A value above 0.5 proves the existence of entanglement \cite{Sackett:2000gm}. 

\noindent\textbf{Polarization stabilization results}. Tomographic measurements of the distributed ion-photon state were interleaved with four polarization stabilization runs: one at the beginning, one at the end and two in between. The results of the polarization stabilization runs show that the fidelity between two consecutively measured processes never dropped below 0.9999. There was therefore no significant change in the polarization dynamics during the tomographic measurement run.

\noindent\textbf{Efficiency and rate.} 
  Over a total of 60 minutes, $k=$ 2167496 attempts were made to deliver a photon to Location B, and $N_\text{d}=$ 2454 detection events were recorded, corresponding to a probability of $P_\text{ip}=1.13(2)\times 10^{-3}$ for delivering an ion-entangled photon, and a photon detection rate of \SI{0.68}{\second ^{-1}}. Of those 60 minutes, 3.2 minutes were used for periodic polarization stabilization, 12.05 minutes for other calibrations, 1.5 minutes for waveplate rotation to change the photon-qubit measurement basis and 7.95 minutes to reprogram our hardware for controlling pulse generation. Thus, only 35.3 minutes were spent running the photon distribution sequence, during which time the photon delivery rate was \SI{1.16}{\second ^{-1}}.
The temporal profile of the detected photons at Location B is shown in Fig.~\ref{fig:ion-photon}(a). This photon wavepacket is presented as a probability density $P_\text{d}(t) = N_\text{d}/(\kappa \times \delta t)$, where $N_\text{d}$ is the number
of detection events registered in a time bin $\delta t=$\SI{1}{\micro\second} and $\kappa$ is the number of attempts to generate a photon. 
To obtain the simulated wavepacket in Fig.~\ref{fig:ion-photon}(a), we numerically solve a master equation describing the photon-generation process, as detailed in Appendix \ref{sims1}. In summary, the model is implemented in a Python package which solves a master equation describing the ion-cavity interaction. It operates in a Hilbert space comprising 18 electronic states in the ion and two frequency-degenerate modes of the cavity \cite{Brandstatter2013}.  All parameters in the model, except the photon detection path efficiency, are independently measured via calibration experiments. 
The total probability of the simulated wavepacket shown in Fig.~\ref{fig:ion-photon}(a) has been normalized to match the measured value $P_\text{ip}=1.13(2)\times 10^{-3}$. 

Table~\ref{table:eff budget} presents an efficiency budget for the expected value $P_\text{ip,calc}$, where each row entry gives the efficiency of a distinct part of the photon path, beginning with the probability $P_{1}$ that the ion emits a photon into the cavity and ending with the telecom detector efficiency $P_{8}$.  
Values P2 through P8 are obtained from independent calibration experiments (see Appendix~\ref{appendix:path_eff}). The probability for generating a cavity photon is obtained from the same numerical simulation used to produce the simulated photon wavepacket in Fig.~\ref{fig:ion-photon}(a). The product of $P_{1}$ through $P_{8}$ yields the total expected maximum probability for detecting a photon at Location B of $P_\text{ip,calc}=1.8(2)\times 10^{-3}$. This is a maximal value, as the values $P_{1}$ through $P_{8}$ are obtained for isolated and optimized individual parts and do not consider additional losses incurred when connecting the parts.  We attribute the lower measured value $P_\text{ip}=1.13(2)\times 10^{-3}$ to these additional losses. The photon path element that contributes the largest loss is Channel Q. 
 
\noindent\textbf{Density matrix, concurrence and fidelity.} 
The absolute value of the tomographically reconstructed ion-photon density matrix $\rho^{\text{B}}_{\text{ip}}$ is presented in Fig.~\ref{fig:ion-photon}(b). The state $\rho^{\text{B}}_{\text{ip}}$ has a Bell state fidelity $F=0.88(2)$ and a 
concurrence $C=0.77(3)$, demonstrating entanglement between the ion and remotely delivered photon. 
To identify the origins of the $1-0.88(2)=0.12(2)$ state infidelity, we develop a model for the ion-photon Bell state fidelity, as detailed in Appendix \ref{appendix:ion-photon infid} and summarized here.  
The model includes two error sources: first, imperfections in the ion-photon state measured before QFC; and second, the detector background counts at Location B. 
The first error source is quantified by switching the photon to the \SI{854}{\nano\meter} photon analysis board at location A and performing ion-photon state tomography, yielding a Bell state fidelity of $F=0.953(7)$. After subtracting photon-detector background counts from the raw tomographic data, we obtain a background-subtracted state with a fidelity $F=0.975(3)$. 
That state is our best estimate of the entangled state injected into the QFC setup and has an infidelity of $1-F=0.025(3)$. 
For the second error source, we extract a background count rate of 2.04(14)~cps from the data set used to produce Fig.~\ref{fig:ion-photon}(b).
Using these two values, the model predicts a remote ion-photon Bell state infidelity of $0.095(8)$, which is consistent with our value of $0.12(2)$ to within one standard deviation of statistical uncertainty. 

Of the 2.04(14)~cps background rate, half originates from noise photons added by the QFC process (Appendix \ref{appendix:setup-location A}), and the rest is due to detector dark counts. The QFC noise is broadband \cite{Krutyanskiy2017} and could be reduced by narrowing the filtering bandwidth without reducing $P_\text{ip}$ (our photons have a bandwidth of $< $~\SI{1}{\mega\hertz}, whilst the narrowest filtering bandwidth in the QFC setup is \SI{0.5}{\giga\hertz}). A further way to mitigate infidelity from background counts is to lower the QFC crystal's phase-matching temperature, which suppresses anti-Stokes Raman-scattered photons \cite{Krutyanskiy2017}. Identifying the causes of infidelities in the ion-photon state before frequency conversion and developing an infidelity budget will require a separate study; we expect multiple physical contributions at the sub-percent level, including imperfections in the laser-driven ion-analysis pulses and in the photon-polarization analysis optics.

\begin{table}[h]
	\caption{Efficiency budget for the probability of generating an ion-entangled photon, transferring it over the deployed fiber and detecting it at Location B, for the experiment presented in Sec.~\ref{ion-photon entanglement}. $P_1$ to $P_8$ are independently estimated losses from eight elements in the photon path.}
	\label{table:eff budget}
	\centering
	\setlength{\tabcolsep}{6pt}
	\renewcommand{\arraystretch}{1.1}
	\begin{tabular}{|l|c|S[table-format=1.4(2)]|}
		\hline 
		\textbf{Element} & \textbf{Symbol} & \textbf{Efficiency}\\ 
		\hline
		Cavity photon generation & $P_1$ & 0.59 \\
		To exit the cavity & $P_2$ & 0.43(3) \\ 
		Path to QFC & $P_3$ & 0.61(3) \\
		QFC & $P_4$ & 0.22(1) \\
		Channel Q & $P_5$ & 0.192(3) \\
		FBG & $P_6$ & 0.54(2) \\
		Photon analysis board & $P_7$ & 0.62(2)\\ 
		Average detector & $P_8$ & 0.82(2) \\
		\hline
		\textbf{Total product of above} & \textbf{\boldmath $P_{\mathrm{ip,calc}}$} & \bfseries 0.0018(2)\\
		\hline
		Total measured & $P_{\mathrm{ip}}$ & 0.0013(2)\\
		\hline
	\end{tabular}
\end{table}

\section{Repeater node demonstration} \label{Repeater node demonstration}

\noindent\textbf{Background.} 
Following the original proposal for a quantum repeater \cite{Briegel1998}, the action of a central repeater node is to first establish bipartite entanglement between itself and its neighboring nodes separately, exploiting its ability to store quantum information in memory, then to perform entanglement swapping to establish entanglement directly between its neighbors. Entanglement swapping is ideally realized via 
a deterministic Bell state measurement (DBSM) on the repeater node's qubits. 

In our network, following the approach of \cite{KrutyanskiyCanteriMeraner2023}, the ion trap functions as a repeater node and the two photon analysis stations (photonic nodes) are the neighboring nodes. Entanglement is first established between one trapped ion and a photon detected in Location B, then between a second, co-trapped ion and a photon detected in Location A. Finally, a combination of a laser-driven quantum logic gate and logical-state readout of the ion qubits realizes the DBSM on the ion qubits. There are four outcomes of the DBSM, each ideally heralding one of four orthogonal Bell states of the distributed photon qubits. 

\noindent\textbf{Setup.} 
The employed experimental setup is shown in Fig.~\ref{fig1_map_and_intro}(c) and differs from the setup in the previous section in that two ions and the fiber-coupled optical switch are employed. The switching time is on the order of a microsecond, allowing sequentially emitted photons to take different paths. The switch has a transmission probability of 0.64(3) to the local analysis board and 0.640(4) to the QFC setup. 
The transmission efficiency of the QFC setup for this experiment was improved to 0.36(4), from the 0.22(1) value in the previous section, as summarized in Appendix \ref{appendix:setup-location A}. The photon detection efficiency budget at Location B is thus the one in Table \ref{table:eff budget} with the addition of the switch transmission factor, a modification of the QFC value to 0.36(4) and new simulated cavity photon generation efficiencies $P_1^1=$ 0.61 and $P_1^2=$ 0.52 for ions 1 and 2, respectively. The calculated value $P_\text{ip,calc}$ is $1.9(3) \times 10^{-3}$. 
The new values for $P_1$ are discussed in Appendix \ref{repeatersims}. 

\noindent\textbf{Repeater node protocol.} 
The protocol has four parts: Initialization, Loop 1, Loop 2 and DBSM. Initialization prepares two ions in $|S\rangle$ and the ground state of the radial center-of-mass motional mode with highest frequency. 
Up to 20 attempts are then made in Loop 1 to deliver a \SI{1550}{\nano\meter} photon, emitted by and entangled with ion 1, to Location B. In the case of detection of that remote photon during any attempt, Loop 1 is aborted, and up to 22 attempts are next made in Loop 2 to deliver an \SI{854}{\nano\meter} photon that is entangled with the co-trapped ion (ion 2) to Location A. During Loop 2, the qubit state encoded in ion 1 is stored in the ground state manifold, and dynamical decoupling is executed. 
In the case of detection of the second photon at Location A, both ion-qubit states are transferred into optical encodings and the DBSM is performed: First, the MS logic gate is applied; Second, the logical state of each ion qubit is measured. 
The repeater protocol is now complete. More details are given in Appendix \ref{repeaterprotocol}. 
The protocol is repeated for 36 polarization settings at the photonic nodes in order to characterize the photonic states via quantum state tomography.

\begin{figure}[t]
	\centering
	\includegraphics[width=\columnwidth]{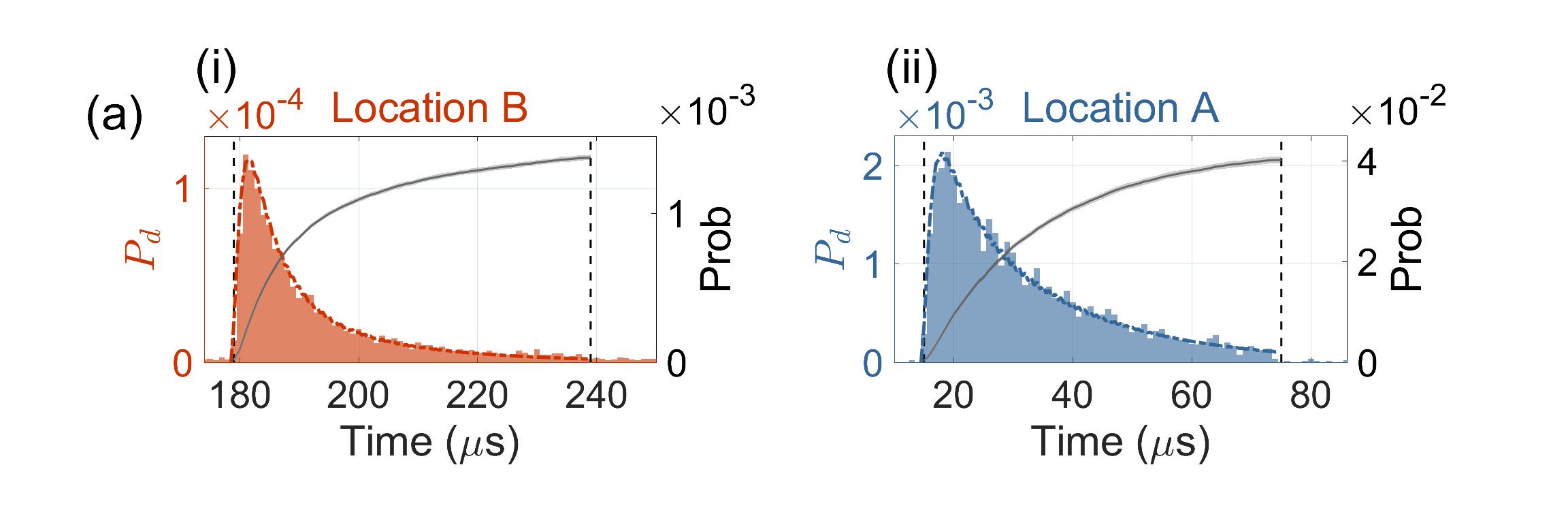}
	\includegraphics[width=\columnwidth]{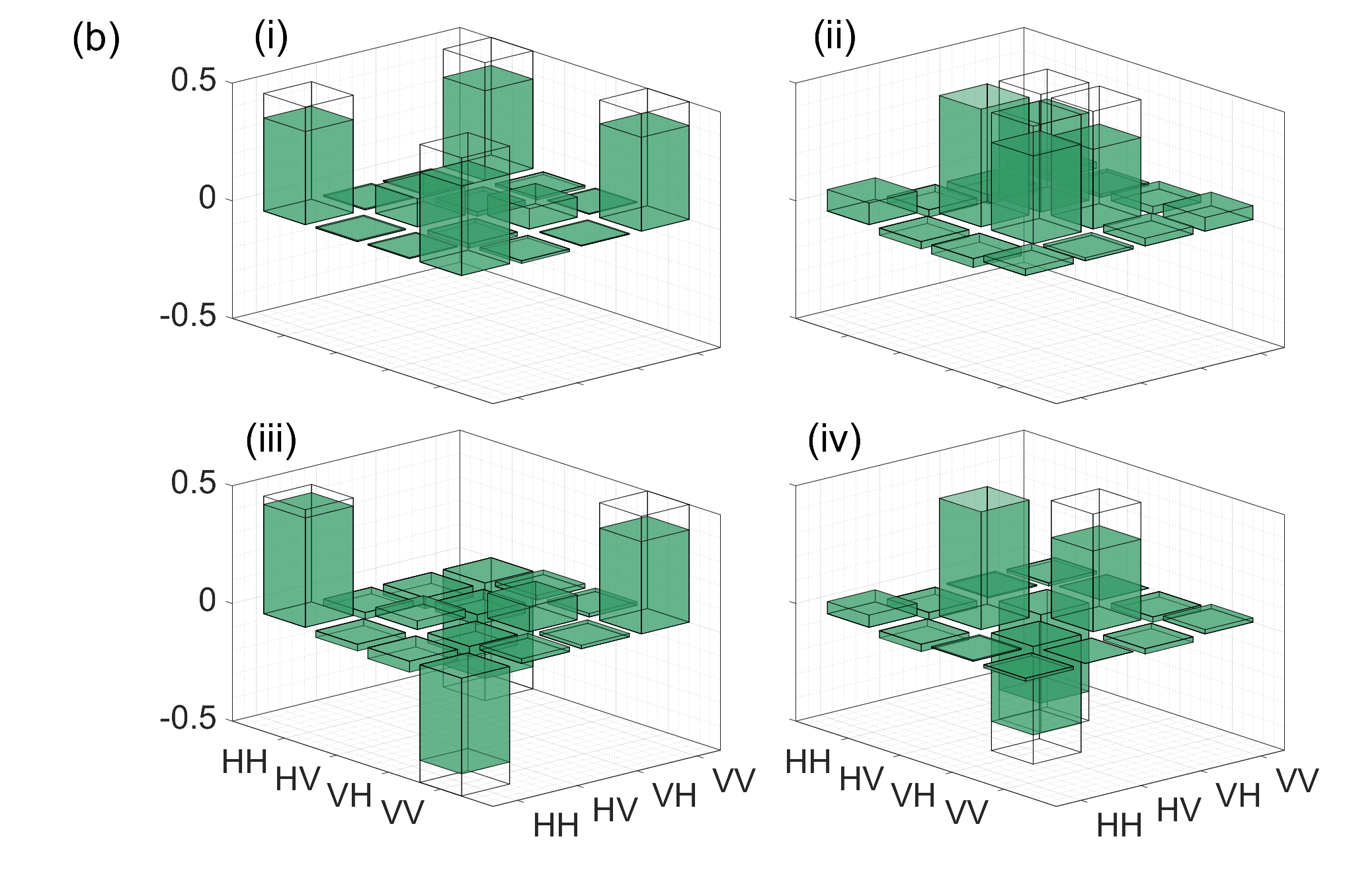}
	\caption{
		\textbf{Repeater node protocol results.} 
		(a) Left axes: probability distribution ($P_d$) for single-photon detection per \SI{1}{\micro\second} time bin. Time origins: Raman laser pulse onset. Colored dashed lines: theoretical model. Dashed black lines: windows in which DBSM is performed. Right axes: cumulative photon detection probability. (b) Colored bars show measured real parts of the density matrix of the two single-photon polarization qubits shared by nodes A and B. Fidelities with targeted Bell states (wireframe) are (i) $78^{+3}_{-3}\%$, (ii) $80^{+3}_{-4}\%$, (iii) $83^{+3}_{-4}\%$, (iv) $84^{+3}_{-3}\%$. Absolute values of the imaginary parts do not exceed 0.06.}
		\label{figure5}
\end{figure}

\noindent\textbf{Protocol differences.}
The protocol differs from the one of \cite{KrutyanskiyCanteriMeraner2023}. 
In \cite{KrutyanskiyCanteriMeraner2023} a \emph{symmetric} protocol was implemented, in which attempts to deliver photons to both detection stations were made in the first loop (Loop 1); whichever ion qubit succeeded in delivering its photon was then stored in memory, and attempts to deliver the final photon from the other ion were made in the second loop (Loop 2). The detection probability at the two detection stations was similar: around $1\times10^{-3}$. 
In the present work, the detection probabilities are strongly asymmetric, with a higher value at the local node. Running the symmetric protocol would almost always lead to detection at the local node first and the need to store the qubit in memory until the remote node succeeds, which would take on average $1\times10^{3}$ attempts). 
We therefore run an asymmetric protocol, which attempts to deliver to the remote node first and the local node second, thereby strongly reducing the memory storage time at negligible cost to the overall success probability. 
Minimizing the time spent in memory, and therefore the number of photon generation attempts in Loop 2, is beneficial not only due to the reduced decoherence of the memory but also because the expected photon generation efficiency in Loop 2 is lower than in Loop 1 due to the absence of optimal electronic state preparation (optical pumping).

\noindent\textbf{Interleaved calibrations}. Execution of the protocol was interleaved with five polarization stabilization runs, spaced by an average of 1.5 hours. The mean fidelity between consecutive fiber process matrices was 0.988(7), where the uncertainty is given by the range. A second interleaved calibration measurement was tomographic reconstruction of the distributed ion-photon states, which was repeated five times spaced by a similar average time. Here, the repeater protocol was executed with one modification: the MS gate was replaced with laser pulses that enabled tomographic analysis of the ion-photon states. Adding the data from those five measurement sets yields two density matrices, one between ion 1 and the \SI{1550}{\nano\meter} photonic polarization qubit delivered remotely ($\rho^{\text{B}}_{\text{ip,rep}}$), and one between ion 2 and the \SI{854}{\nano\meter} photonic polarization qubit delivered locally ($\rho^{\text{A}}_{\text{ip,rep}}$). Those matrices represent our best estimates of the distributed ion-photon states over the time period during which the repeater node protocol was run.

\noindent\textbf{Probabilities and memory advantages.} The repeater protocol was executed 67000 times over 4.5 hours, during which 4479061 attempts were made to establish photonic entanglement between nodes A and B, and 3369 successes occurred. 
After subtracting the time spent performing calibrations, rotating waveplates and reprogramming, 81 minutes were spent running the repeater node sequence.  The success rate is therefore \SI{0.69}{\second^{-1}}, and the success probability is  $P_\text{s} = 7.52(13)\times10^{-4}$. Successes correspond to asynchronous detection of a photon at each node, where each photon is detected within a 60 $\mu$s time window containing the photon wavepackets (Fig.~\ref{figure5}(a)).  
The value of $P_\text{s}$ can be compared with the expected 
probability $P^{\text{single}}_\text{s}$ to establish Location A-B photon-photon entanglement without exploiting the ion memory.
We calculate $P^{\text{single}}_\text{s} = P_{\text{B},1}\times P_{\text{A},2}=5.55(13)\cdot 10^{-5}$, where $P_{\text{B},1}=1.38(2)\cdot 10^{-3}$ and $P_{\text{A},2}=4.02(7)\cdot 10^{-2}$ are the measured probabilities for photon detection at Locations B and A, respectively. 
The ion memory thus increases the success probability by a factor $\alpha = P_\text{s}/P^{\text{single}}_\text{s} = 13.6(4)$.
In principle, by performing enough Loop 2 attempts, one can approach unit probability for the detection of the second photon, yielding the maximum memory factor for our experiment of $\alpha^{\text{max}} = 1/P_{\text{A},2} = 24.9(4)$.  
In our experiment, the 22 attempts in Loop 2 resulted in a probability for detecting the second photon of 0.556(9), limiting $\alpha$ to be $\alpha^{\text{max}}\times 0.556(9)$. 
Thus, increasing the number of Loop 2 attempts could increase the success probability only by less than a factor of two.  

\noindent\textbf{Photon-photon states.} Fig.~\ref{figure5}(b) presents the four two-photon density matrices reconstructed from the 3369 total successes, and each measured conditioned on a different logical state of the ion-qubits.
As in Section III, we numerically search for the single-qubit rotations that maximize the fidelity with Bell states. Here, the same two single-qubit rotations are applied to all four matrices, and the search maximizes the mean fidelity over the matrices. 
The Bell state fidelities and concurrences of the measured states are $[0.78^{+3}_{-3}, 0.80^{+3}_{-4}, 0.83^{+3}_{-4}, 0.84^{+3}_{-3}]$ and $[0.59^{+8}_{-5}, 0.63^{+6}_{-7}, 0.72^{+6}_{-8}, 0.71^{+6}_{-6}]$, proving entanglement. 
Our repeater protocol thus succeeds in establishing entanglement between photons detected at the two analysis stations.  

Using the data underlying all four photon-photon states, we numerically simulate the conditional local rotations (feedforward) that would deliver a single Bell-like state to the remote nodes, as in quantum teleportation \cite{BennettBrassardCrepeauJozsaPeresWootters1993}. This calculation is performed as described in Ref.~\cite{KrutyanskiyCanteriMeraner2023} (Section 2.C, Supplementary Material) and yields 
the state $\rho_{\text{pp}}^{\text{ff}}$ with Bell state fidelity $81.6^{+1.9}_{-1.4}\%$. If we were able to implement perfect feedforward to the photon-photon states, this is the fidelity of the state that would be delivered.

\noindent\textbf{Protocol fidelity limits.}
To identify the origins of protocol infidelity, we develop a simple model of the distributed photon-photon state $\rho_{\text{pp}}^{\text{ff}}$ that considers only imperfections in the measured ion-photon states. The model takes as input the ion-photon states $\rho^{\text{B}}_{\text{ip,rep}}$ and $\rho^{\text{A}}_{\text{ip,rep}}$, obtained during the interleaved calibrations. Perfect DBSM and feedforward operations are then applied numerically to these states, yielding a predicted photon-photon Bell-like state. The model predicts a photon-photon Bell-state fidelity of $82^{+2}_{-2}\%$, which is statistically consistent with the experimental  value of $81.6^{+1.9}_{-1.4}\%$. Therefore, 
to within the statistical precision of our measurements, the infidelities observed in $\rho_{\text{pp}}^{\text{ff}}$ are already present in $\rho^{\text{B}}_{\text{ip,rep}}$ and $\rho^{\text{A}}_{\text{ip,rep}}$, distributed in the repeater node protocol.
These ion-photon states have fidelities with maximally entangled states of 0.87(3) ($\rho^{\text{B}}_{\text{ip,rep}}$) and 0.91(3) ($\rho^{\text{A}}_{\text{ip,rep}}$). 

The origins of infidelity in the ion-photon states are now discussed, beginning with $\rho^{\text{B}}_{\text{ip,rep}}$. The Bell-state infidelity of $\rho^{\text{B}}_{\text{ip,rep}}$ is $1-0.87(3)=0.13(3)$.
Table~\ref{table:remote ionphoton infid budget} summarizes four contributions to the infidelity of $\rho^{\text{B}}_{\text{ip,rep}}$, ordered from largest to smallest. 
The contribution due to background counts is $0.07(1)$, corresponding to the infidelity introduced into an otherwise perfect Bell state $\ket{\psi(0)}$ by the $2.06(18)$ cps background measured at Location B (Appendix~\ref{appendix:ion-photon infid}).
The contribution due to decoherence of the ion 1 memory is 0.060(9), as calculated from characterizations of the ion-photon Bell state fidelity before and after memory storage in Location A (Appendix \ref{appendix:memorycost}). 
The characterization before memory storage also yields the contribution of 0.032(5) due to imperfections in the initial ion-photon entangled state produced at ion 1; here, background counts are subtracted, 
Finally, the contribution due to drifts in fiber birefringence is 0.014(6), as calculated from the fiber process matrices measured during interleaved stabilisation runs (Appendix~\ref{appendix:repeater_ip_infid_pol_drift}).
The sum of all contributions is $0.176(16)$, which is statistically consistent with the experimental infidelity of $0.13(3)$ to within two standard deviations of the combined uncertainty (Table~\ref{table:remote ionphoton infid budget}).

\begin{table}[ht]
	\centering
	\begin{tabular}{|l|S[table-format=1.3(2)]|}
		\hline
		Error source & {Infidelity} \\
		\hline
		A. Background counts & 0.07(1) \\
		B. Ion-memory & 0.060(9) \\
		C. Initial ion-photon state & 0.032(5) \\
		D. Fiber rotations & 0.014(6) \\
		\hline
		\textbf{Total above} & \bfseries 0.176(16) \\
		\hline
		Total measured & 0.13(3) \\
		\hline
		\textbf{Difference} & \bfseries 0.046(34) \\
		\hline
	\end{tabular}
	\caption{Infidelity budget for the ion-photon entangled state ($\rho^{\text{B}}_{\text{ip,rep}}$) distributed to Location B,  during the repeater node demonstration of Section~\ref{Repeater node demonstration}.}
	\label{table:remote ionphoton infid budget}
\end{table}

The origins of ion-memory imperfections are now discussed. Magnetic-field fluctuations at the power-line frequency of \SI{50}{\hertz} modulate the ground-state qubit transition frequency via the linear Zeeman effect. In our system, a \SI{50}{\hertz} noise amplitude of \SI{36(3)}{\micro\gauss} was previously measured. A model for the ground-state qubit under our dynamical decoupling sequence (Appendix \ref{appendix:modelSpinecho}) predicts that this amplitude would reduce the fidelity of an otherwise perfect ion-photon state by 0.03; a noise amplitude of \SI{54}{\micro\gauss} would be required to produce a 0.06 reduction. Because the \SI{50}{\hertz} noise amplitude can vary by tens of microgauss from day to day in our system and was not measured on the day the repeater-node data were taken, \SI{50}{\hertz} noise cannot be excluded as the sole cause of the observed 0.06 fidelity decrease due to memory storage. We do not expect any significant detrimental effects of Raman laser cross-talk on the ion-memory during the protocol. 

That state $\rho^{\text{A}}_{\text{ip,rep}}$ is not stored in memory, and the detector background counts contribute a fidelity drop of only 0.007(4). The 0.09(3) state infidelity is higher than the value reported in Sec. \ref{ion-photon entanglement}, and than we have previously achieved with multiple ions under comparable conditions \cite{KrutyanskiyCanteriMeraner2023, KrutyanskiyCanteriMeranerKrcmarskyLanyon2024, CanteriBateMishraFriisKrutyanskiyLanyon2025}.   
We have not yet identified the origins of the missing fidelity.  
To within statistical precision, the Bell state fidelity of $\rho^{\text{A}}_{\text{ip,rep}}$ is limited by its purity; thus, we are searching for error sources corresponding to non-unitary transformations of the ion-photon state.

\section{Conclusion and outlook} 
\label{Conclusion_and_outlook}

We verified the operation of a trapped-ion QRN with one of its fiber segments deployed across a city, employing the preexisting fiber infrastructure. By combining cavity photon generation, storage in ion-qubit memories, telecom-wavelength photon conversion and a deterministic Bell-state measurement, the repeater node established entanglement between the polarization of asynchronously generated photons across the city with a Bell-state fidelity of $81.6^{+1.9}_{-1.4}\%$ and a two-photon rate of \SI{0.69}{\second^{-1}}.  

The deployed fiber did not significantly contribute to the state infidelity. 
It was possible to mitigate the slow drifts in fiber birefringence by applying periodic corrections and to mitigate the photon noise coupled along the fiber length by employing a GHz-bandwidth filter at the output.
Compared with a standard spooled fiber, the deployed fiber had about half the transmission due to its numerous splices and connectors and accordingly achieved roughly half the rate.

Instead, the fidelity and rate were limited by imperfections in the repeater node and detection stations. 
The main contributions to the entangled-state infidelity were from dark counts at the remote photon detectors (1-2 cps), noise photons produced by the QFC setup, imperfections in the prepared ion-photon state prior to frequency conversion and ion-qubit decoherence from magnetic-field noise. Significant fidelity improvements are expected by identifying and mitigating remaining ion-photon state preparation errors and exploiting more advanced spin-echo sequences to extend ion coherence.  

The two-photon rate was limited by the probability to deliver and detect a photon at the remote station $P_{\text{B},1}=1.38(2)\cdot 10^{-3}$, due to the photon  loss processes between and including the QRN and the remote station. At least an order of magnitude improvement should be straightforward to achieve ($P_{\text{B},1}=1\cdot 10^{-2}$) by combining, e.g.,  the ion-cavity performance of \cite{KrutyanskiyCanteriMeraner2023} ($P^1_1 P_2=0.940(2)\times0.78(2)=0.733(19)$), the QFC conversion efficiency of \cite{PhysRevLett.124.010510} ($P_4=0.57$) and readily available lower-loss filters ($P_6=0.9$). 

An important next step is to use the QRN to establish stored entanglement between distant ions. 
For example, with the central improved QRN remaining in the lab, consider deploying 
two detection stations and two duplicate improved QRNs ($P_{\text{B},1}=1\cdot 10^{-2}$) in opposite directions, across and outside of the city, to form a  prototype repeater-node chain over a fiber distance of \SI{68}{\kilo\meter} ($4 ~\times$ \SI{17}{\kilo\meter} fiber segments). 
When using a two-photon click scheme  \cite{Duan03, Feng03, Simon03, DuanMonroe2010, KrutyanskiyGalliKrcmarsky2023}, we calculate that the expected mean time to herald stored entanglement between an ion in each of the two deployed QRNs is \SI{5.4}{\second}, requiring a mean ion-qubit coherence time of \SI{3.6}{\second}. 
Significantly shorter times could be obtained using a single-photon click scheme \cite{PhysRevA.59.1025,PhysRevLett.83.5158}, at the expense of additional phase-stability requirements, or by implementing a multi-ion approach \cite{KrutyanskiyCanteriMeranerKrcmarskyLanyon2024, PhysRevLett.98.190503, PhysRevLett.98.060502}. 
With multi-second coherence times recently achieved in our system~\cite{Bate2026}, prototype metropolitan-scale QRN chains are already within reach. Details on the calculations are given in Appendix \ref{appendix:rates_distant}.

QRN-assisted entanglement distribution provides a route to quantum networks that span continental distances, with higher rates than direct transmission \cite{KrutyanskiyCanteriMeraner2023, Avis:2023im, Weinfurter2020_theory_repeater}. As detailed in \cite{KrutyanskiyCanteriMeraner2023}, the  requirements for our QRNs to enable ion-ion entanglement over \SI{800}{\kilo\meter}  at hertz rates are feasible with a concerted research effort. 
The results presented in this manuscript show that future long-distance quantum networks could exploit the existing telecom-wavelength fiber infrastructure, eliminating the need to build a dedicated fiber network. 

\section*{Data Availability}
The data underlying the results in this manuscript will be made publicly available on Zenodo upon publication in a peer-reviewed journal.

\begin{acknowledgments}
This work was funded in part by: the European Union under the DIGITAL-2021-QCI-01 Digital European Programme under Project No. 101091642 (Project name `QCI-CAT'), and the European Union's Horizon Europe research and innovation programme under grant agreement No. 101102140 (Project name `QIA-Phase 1'); the \"Osterreichische Nationalstiftung f\"ur Forschung, Technologie und Entwicklung (Project name `AQUnet'); 
the Austrian Science Fund (FWF) 
[Grant DOIs: 10.55776/PAT1502824, 
10.55776/F71, 
10.55776/P34055 
 and 10.55776/COE1]. 
V.K. acknowledges support from a Quantum Science Austria postdoctoral fellowship. 
The opinions expressed in this document reflect only the authors' views and reflects in no way the European Commission's opinions. The European Commission is not responsible for any use that may be made of the information it contains. For open access purposes, the author has applied a CC BY public copyright license to any author-accepted manuscript version arising from this submission. 

F.K., V.K., A.W., F.B. and T.R. took the experimental data. 
F.K., A.W., T.R., V.K. and F.B. developed the experimental setup, with key contributions from J.H., J.B. and M.C. 
F.K., V.K., T.R., A.W. and B.P.L. carried out the data analysis and interpretation. 
F.K., V.K., T.R. and F.B. performed the modelling. 
B.P.L. and F.K. primarily wrote the manuscript, with key contributions from A.W., V.K., T.E.N., T.R. and F.B.
The project was supervised by B.P.L.
\end{acknowledgments}

\appendix

\renewcommand{\thesubsection}{\Alph{section}\arabic{subsection}}

\section{Experimental setup and methods}\label{appendix:experimental setup}

\begin{figure*}[!t]
\centering
\includegraphics[width=\textwidth]{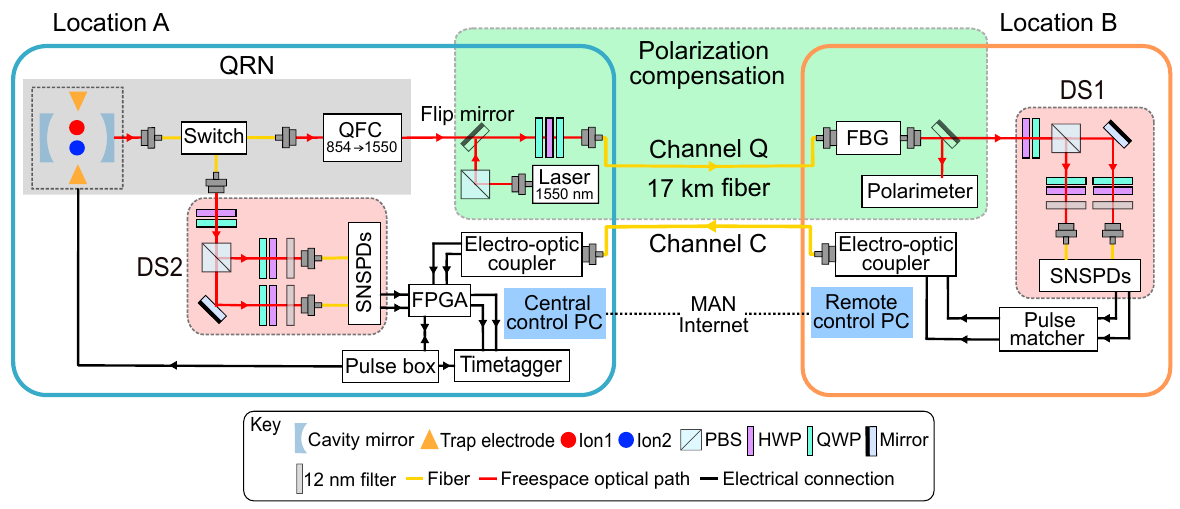}
\caption
{\textbf{Detailed schematic of the two-location quantum network}.
The two locations are connected by two \SI{17}{\kilo\meter} single-mode (SMF-28) fibers. Channel Q transports single photons from Location A to Location B, whereas Channel C returns detector signals from Location B to Location A via electro-optic couplers. 
QRN: quantum repeater node; QFC: quantum frequency conversion; DS1(2): detection station 1(2); FBG: fiber Bragg grating. 
The timetagger at Location A records photon-detection timestamps. The Pulse box provides central control of the experimental sequence, including ion-trap control and device triggering; Superconducting nanowire single photon detector (SNSPD) signals are routed to the pulse box via an FPGA for ion-state readout. The green shaded area indicates the polarization-compensation system for Channel Q: at Location A, a laser, polarizer, motorized flip mirror, and motorized waveplates prepare the light before transmission through Channel Q; at Location B, a second motorized flip mirror directs the light to a polarimeter. Computers at the two locations communicate via the University of Innsbruck's metropolitan-area network (MAN) and the Internet.}
\label{figA1_setup}
\end{figure*}

In this appendix section, the main parts of the two-location quantum network are presented in detail: Location A, Location B, Channels Q and C and the polarization stabilization setup. A detailed schematic of the two-location quantum network is provided in Fig. \ref{figA1_setup}. The main parts are indicated in the figure. 

\subsection{Location A}\label{appendix:setup-location A}

\noindent\textbf{Ion trap and cavity.} 
We use a linear Paul blade trap with an integrated optical cavity for photon collection at \SI{854}{\nano\meter}.  For details on the system design, construction and initial cavity characterization see Ref.~\cite{SchuppPhDthesis2021,SchuppKrcmarskyKrutyanskiyMeranerNorthupLanyon2021}. The cavity finesse decreased by a factor of 1.8 relative to the initial value, as explained in the supplemental material of Ref.~\cite{Canteri2025}. The cavity axis is near-perpendicular to the ion-string axis, with a relative angle of $85.9^\circ$. The relevant \SI{854}{\nano\meter} vacuum mode of the \SI{19,906(3)}{\milli\meter}-long near-concentric cavity has a \SI{12.31(8)}{\micro\meter} waist. In the present work, up to two \ch{^{40}Ca^+} ions are trapped and positioned near the neighboring anti-nodes of the cavity standing wave. The procedure used for positioning the ions with respect to the cavity and the focused laser is described in supplemental material of Ref.~\cite{KrutyanskiyCanteriMeraner2023}. 
A single cavity photon is generated from an ion of choice via a bichromatic cavity mediated Raman transition (BCMRT) driven by a single-ion-focused \SI{393}{\nano\meter} Raman laser beam \cite{Stute2012}, following the procedure described in the supplemental material of Ref.~\cite{KrutyanskiyCanteriMeraner2023}.\\

\begin{figure}[t]
	\includegraphics[width=\linewidth]{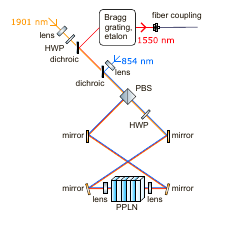}
	\caption{
		\textbf{QFC setup for difference frequency generation.} We implemented a polarization-preserving setup in the Sagnac-configuration \cite{Bock2018}.  
		HWP: half-wave plate, PBS: polarizing beam splitter, PPLN: periodically-poled lithium niobate. } 
		\label{fig:detailledQFCsetup}
\end{figure}

\noindent\textbf{QFC setup and characterization.}\label{appendix:QFC Setups and Characterization}
For fiber transmission over tens of kilometers, \SI{854}{\nano\meter}  photons from a trapped \ch{^{40}Ca^+} ion are quantum-frequency converted to \SI{1550}{\nano\meter} in a \SI{47}{\milli\meter} long PPLN ridge waveguide (NTT Electronics) at a phase-matching temperature of 27\;$^{\circ}$C using a \SI{1901}{\nano\meter} pump laser. We implemented a quantum frequency conversion (QFC) setup in the Sagnac configuration \cite{Bock2018}. Fig.~\ref{fig:detailledQFCsetup} shows the optical layout. The crystal is oriented for conversion of vertically polarized (V) light. The 1901\;nm pump light passes a half-wave plate to balance power and thus conversion efficiency in both paths. Incoming \SI{854}{\nano\meter} photons and the \SI{1901}{\nano\meter} light are split at a PBS into V and horizontal (H) polarization components; the H arm includes a half-wave plate that rotates H to V for both \SI{854}{\nano\meter} and \SI{1901}{\nano\meter}. Both paths are coupled into the PPLN waveguide, where conversion to \SI{1550}{\nano\meter} occurs. After outcoupling, the half-wave plate in the H arm flips the polarization of all three wavelengths from V to H. Converted photons are filtered by: (i) a dichroic mirror with a \SI{0.2}{\micro\meter} bandwidth and $>0.99$ reflection at \SI{1550}{\nano\meter} to separate \SI{1550}{\nano\meter}  from \SI{854}{\nano\meter} and \SI{1901}{\nano\meter}; (ii) a volume Bragg grating (VBG) with a \SI{10}{\giga\hertz} bandwidth and $>0.99$ diffraction efficiency; and (iii) a fused silica etalon with a \SI{0.5}{\giga\hertz} bandwidth and up to 0.95(1) transmission efficiency. The \SI{1550}{\nano\meter} photons are finally fiber-coupled with an efficiency of 0.83(2) after the filter stage. \\ 
The fiber-coupled noise photon rate, when only sending pump light into the QFC setup, is 3.6(1)\;cps per \SI{100}{\milli\watt} of \SI{1901}{\nano\meter} pump power, where the reported pump power is measured before the lens in the top-left corner of Fig.~\ref{fig:detailledQFCsetup}, at the \SI{1901}{\nano\meter} input. This noise rate is used in subsequent calculations and was determined with an optimised (clean) etalon filter. 
The \SI{1901}{\nano\meter}  power is approximately equally split between the two paths of the interferometer. 
The external conversion efficiency ($P_4$, Table \ref{table:eff budget}) is the probability that an \SI{854}{\nano\meter} photon entering the setup is converted to \SI{1550}{\nano\meter} and exits the setup coupled into the output fiber. For an input pump power of \SI{0.73(1)}{\watt}, the maximum external conversion efficiency that has been achieved in the system is $P_4^{\mathcal{max}}{=}~$0.42(1) and shows no statistically significant dependence on input polarization. For the experiments presented in this work, lower external conversion efficiencies were achieved, as now described. 

In the experiment distributing ion-photon entanglement from a single trapped ion, presented in Sec. \ref{ion-photon entanglement}, the external conversion efficiency of the QFC setup was $P_4^{\mathcal{max}}=~$0.22(1) for a pump power of \SI{0.61(1)}{\watt}. There are two reasons for the lower external conversion efficiency. First, a lower pump power was used due to concerns of damaging the crystal with higher powers at that time. Second, the etalon transmission efficiency was only 0.70(1) instead of 0.95(1) due to some contamination on its surfaces. After correcting for the lower etalon efficiency, the total expected fiber-coupled noise photon rate at the output of the QFC setup, for \SI{0.61(1)}{\watt} of input pump power, is 16.3(6)\;cps.
The expected rate at which these noise photons would be detected at the remote Location B, averaged over the two detectors, is $P_5P_6P_7P_8\times$16.3(6)$~{=}~$0.9(1)\;cps, where the values from Table \ref{table:eff budget} have been used.

In the repeater node experiment, presented in Sec. \ref{Repeater node demonstration}, the external conversion efficiency of the QFC setup had improved to $P_4^{\mathcal{max}}{=}~$0.36(4), for a pump power of \SI{0.73(1)}{\watt}. The increase in efficiency resulted from cleaning the etalon. The value is still below the best achieved, which we attribute to multimoding of the \SI{1901}{\nano\meter} pump laser. Such multimoding lowers the conversion efficiency and introduces conversion efficiency fluctuations. The total fiber-coupled noise photon rate at the output of the QFC setup, for \SI{0.73(1)}{\watt} of input pump, is expected to be 26.3(8)\;cps. The expected rate at which these noise photons would be detected at the remote Location B, averaged over the two detectors, is $P_5P_6P_7P_8\times$26.3(8)\;cps$~{=}~$1.4(2)\;cps, where values from Table \ref{table:eff budget} have been used. \\

\noindent\textbf{Fiber switch.} After the photon exits the ion-cavity through the output mirror, it is coupled into a single-mode fiber-coupled optical switch and sent either to a polarization analysis setup or the QFC setup (Fig.~\ref{fig1_map_and_intro}(b)). 
We use an Agiltron polarization-preserving switch (model NPSW-128113333-SP) driven at \SI{100}{\kilo\hertz}, with single-mode fiber on both ports. The photons are coupled into the input fiber of the switch and can be switched using a TTL signal to either of two fiber outputs. We measure total in-to-out photon transmission efficiencies of 0.64(3) (sent to Location A) and 0.640(4) (sent to Location B) with cross-talks of 0.03(2) and 0.0048(4) respectively.\\

\noindent\textbf{SNSPDs.}
The two detectors used at Location A are superconducting nanowire single-photon detectors (SNSPD) from the company Scontel LLC. The measured efficiencies of both detectors at \SI{854}{\nano\meter} are 0.82(5). The dark count rates are in the range of 1-2\;cps (sum over both detectors), obtained by repeated measurement on different days with caps applied to the input fibers.

\subsection{Location B}\label{appendix:setup-location B}

\noindent\textbf{FBG.} 
The fiber Bragg grating (FBG) is a commercially available optical filter from AOS GmbH with a center wavelength of \SI{1551.370}{\nano\meter} in vacuum. The filter has an adjustable wavelength tuning range of $\pm \SI{0.24}{\nano\meter} (\pm \SI{30.0}{\giga\hertz})$ and a full-width at half-maximum (FWHM) bandwidth of \SI{14(2)}{\pico\meter}, corresponding to \SI{1.7(3)}{\giga\hertz}. We measured a transmission efficiency of $0.54$. 

\noindent\textbf{SNSPDs.} The two SNSPDs employed at remote Location B are from the company Single Quantum and are optimized for a detection wavelength 
of  \SI{1546}{\nano\meter}. The detector efficiencies measured by the manufacturer before shipping to Location B were 0.81(3) and 0.83(3). We take the average of these values as our best estimate of the average detector efficiency. Onsite dark count rates measured at the time of installation were 0.82(5) cps and 0.72(5) cps, respectively. All these parameters are for a system optimized at a detection wavelength of \SI{1546}{\nano\meter} on both detectors. Repeated measurements of dark counts on subsequent days, with caps applied to the fiber input connectors on the cryostat, yield total rates (sum over both detectors) that varied in the range of 1-2 cps. \\

\noindent\textbf{Electro-optic coupler.}
The electro-optic coupler is a commercially purchased AFL-4800 single-mode fiber optic link operating at \SI{1550}{\nano\meter}. It has two separate modules, one at Location A and one at Location B. A TTL signal is coupled into the module at Location B via one of the two $\pm~\SI{4.8}{\volt}$ TTL inputs and sent as an optical signal at \SI{1550}{\nano\meter} over Channel C to the module at Location A. Upon receiving the optical signal, the module at Location A generates a TTL signal at one of its outputs. \\

\noindent\textbf{Pulse matcher.}
The outputs of the SNSPDs in Location B are analog pulses with amplitudes \SI{840}{\milli\volt} and \SI{664}{\milli\volt} with duration (FWHM) $<$ \SI{20}{\nano\second} and rise time \SI{0.3}{\nano\second}. To ensure reliable transmission and stable timing of the short pulses with sub-nanosecond edges from the detectors we used a homemade pulse matcher, a pulse-conditioning module (comparator + pulse stretcher) directly after each detector. The conditioned pulses were then fed into one of the electro-optic coupler inputs at Location B and transmitted over Channel C to Location A.\\

\subsection{Fiber Channels Q and C}\label{appendix:setup-Channel}

Channels Q and C each comprise multiple SMF-28 non-polarization-maintaining dark fibers sections together. Characterization of the length and loss of the channels was performed using an optical time-domain reflectometer (OTDR) in three separate sections. \\

\noindent\textbf{Section in Location A.} In this section, the fibers of each channel start in the laboratory on the first floor at Technikerstr.~25 and run in a fiber bundle to the University Central IT Services building, located on the ground floor at Technikerstr.~23. This section was installed and characterized by the Obrist GmbH. According to the company's OTDR report, the fibers are \SI{261.8}{\meter} long, with losses of \SI{0.143}{\mathrm{dB}} (Channel Q) and \SI{0.178}{\mathrm{dB}} (Channel C) at \SI{1550}{\nano\meter}.\\

\noindent\textbf{Section between Location A and Location B.}
The fibers in this section are rented from and characterized by Innsbruck Municipal Utilities (IKB). They start at the University Central IT Services building and run underground, bundled with other city fibers, through the city of Innsbruck to the basement of the university library storage building in the village of G\"arberbach, Mutters. Each channel section consists of 28 dark fibers joined via E2000HLR FC/PC connectors, or by splicing, resulting in a total length of \SI{16.389}{\kilo\meter}. According to the OTDR reports provided by IKB, the losses at \SI{1550}{\nano\meter} for the sections of Channel Q and Channel C are \SI{5.749}{\mathrm{dB}} and \SI{6.316}{\mathrm{dB}}, respectively.\\

\noindent\textbf{Section in Location B.} This section connects the basement of the university library storage building to an office on the first floor, where the photon polarization analysis setup and the SNSPDs are located. This section was installed and characterized by the Obrist GmbH. According to the company's OTDR report, the fibers are \SI{92.3}{\meter} long, with losses of \SI{0.322}{\mathrm{dB}} for Channel Q and \SI{0.106}{\mathrm{dB}} for Channel C at \SI{1550}{\nano\meter}.\\

\noindent The three sections are connected to form two continuous fiber paths, Channels Q and C, spanning the distance between the laboratory in Location A and the photon-detection station in Location B. Adding the results of the OTDR reports, Channels Q and C have a total length of \SI{16.743}{\kilo\meter}, and a total loss of \SI{6.214}{\mathrm{dB}} and \SI{6.600}{\mathrm{dB}}, respectively. The corresponding transmission probabilities of a \SI{1550}{\nano\meter} photon are 0.239 (Channel Q) and 0.219 (Channel C). 

After joining the three sections, the total path transmissions of Channels Q and C were characterized by injecting \SI{1550}{\nano\meter} laser light through their length of \SI{16.743}{\kilo\meter}. In this setup, the fiber-coupled output of a DFB benchtop laser (Thorlabs S3FC1550) was connected to the input of Channels Q/C using a fiber joiner at Location A, and the output of Channels Q/C at Location B was connected to a power meter.
The laser power was measured at both the input and output with Thorlabs S148C power meters. The transmissions, i.e., the ratio of output to input power, are 0.192(3) and 0.205(3). We attribute the difference between the sum of the OTDR-measured losses and these values to the additional joiners needed both to connect the fiber-coupled output of the laser to the first section and to interconnect the three sections.

\subsection{Polarization stabilization}\label{appendix:setup-polstab}

\noindent\textbf{Experimental setup.}
The setup used to characterize and stabilize the polarization dynamics in Channel Q is divided between Location A and Location B, as indicated in green in Fig.~\ref{figA1_setup}. In Location A, the system includes a \SI{1550}{\nano\meter} laser source, a polarizer, a flip mirror, and a set of waveplates. Initially, the laser source was a DFB benchtop laser (Thorlabs S3FC1550). After the FBG was installed at the end of Channel Q in Location B, the laser was replaced by a diode laser (Toptica grating stabilized tunable diode laser with DLC pro digital controller), offering a broader tuning range allowing to match the center frequency of the FBG. 
The light emitted from the laser is polarized using a Glan-laser polarizer (Thorlabs GL10). A motorized flip mirror (Newport 8892-M-K motorized flipper mount) then directs the polarized light into the optical path that is otherwise used for photon transmission. Following the flip mirror, the beam passes through a set of three waveplates arranged in a QWP---HWP---QWP combination that ideally enables the implementation of any unitary operation on the polarization state of light. The light is subsequently coupled into Channel Q.
In Location B, the light is coupled out of Channel Q and extracted from the photon path using a second flip mirror (Newport 8892-M-K motorized flipper mount). The polarization state of the light is then measured with a polarimeter (Thorlabs PAX1000IR2/M). 

\noindent\textbf{Methods.}
To characterize the polarization dynamics in Channel Q, several polarization states are prepared, transmitted sequentially through Channel Q, and measured using the polarimeter. To prepare the polarization states, the optical axis of the first QWP is aligned with the polarizer's output polarization such that it does not alter the light's polarization. The optical axes of the HWP and the second QWP are then adjusted to generate the following polarization states: horizontal, anti-diagonal, right-circular, vertical, left-circular, and diagonal.
The measured output polarization states are used to perform process tomography~\cite{HradilEtAl2004}, thereby obtaining the quantum process that describes how Channel Q transforms the polarization of the transmitted light. The process is represented by the Choi matrix $E(t)$ at time $t$. The process does not necessarily have to be unitary.
The set of three waveplates (QWP---HWP---QWP) is subsequently used to implement a unitary operation intended to invert the unitary part of the process implemented by Channel Q. Specifically, the Choi matrix $E(t)$ is first decomposed into a set of Kraus matrices by using eigenvalue decomposition. 
The Kraus operator associated with the largest eigenvalue of the Choi matrix represents the dominant contribution to the process. The normalized form of this Kraus operator is found to be close to a unitary operator and is denoted as $U(t)$.
Second, the inverse of $U(t)$, denoted as $U^{-1}(t)$, is implemented by the set of three waveplates (QWP---HWP---QWP) such that the waveplates and Channel Q together ideally implement the identity operation. 
The unitary operation implemented by the three waveplates is written as $U(\alpha_1, \alpha_2, \alpha_3) = U_\mathrm{QWP}(\alpha_1) U_\mathrm{HWP}(\alpha_2) U_\mathrm{QWP}(\alpha_3)$, where $U_\mathrm{QWP}(\alpha_i)$ denotes the unitary implemented by one of the QWPs set to an angle $\alpha_i$ and $U_\mathrm{HWP}(\alpha_2)$ denotes the unitary implemented by the HWP set to an angle $\alpha_2$. The difference between  $U(\alpha_1, \alpha_2, \alpha_3)$ and $U^{-1}(t)$ is denoted as $U_\mathrm{diff}$. A numerical search is applied to optimize the parameters $\alpha_1$, $\alpha_2$, $\alpha_3$ (before implementation in the lab) so that the three absolute squares of the matrix elements $|U_\mathrm{diff}^{11}|^2$, $|U_\mathrm{diff}^{12}|^2$, and $|U_\mathrm{diff}^{22}|^2$ are each minimized. The set of waveplates are rotated to the found angles $\alpha_1$, $\alpha_2$, $\alpha_3$ to correct for the change in polarization induced by Channel Q.

\section{Distributed ion-photon entanglement details}\label{appendix:ion-photon}

In this appendix section, further details are provided on our experiment in which a single photon was distributed to the remote Location B, entangled with a single ion in the trap in Location A. The results of that experiment---tomographic-reconstruction ion-photon state---were presented in Fig.~\ref{fig:ion-photon} and first discussed in Sec. \ref{ion-photon entanglement}. 

\subsection{Pulse sequence}\label{appendix:ion-photon pulse seq}
As summarized in the main text, photons are generated and entangled with the ions via a bichromatic cavity-mediated Raman transition (BCMRT), driven by a single-ion-addressed 393 nm Raman laser \cite{Stute2012}. An ion positioned in an antinode of the cavity vacuum field and in the focus of the laser beam emits a photon into the 854 nm vacuum cavity mode. The polarization of that photon is entangled with the final electronic state of the ion that emitted the photon.
The protocol used for ion-photon entanglement generation and characterization consists of three steps: 1. Initialization; 2. BCMRT; 3. Measurement. \\

\noindent\textbf{Initialization.}
Initialization begins with \SI{2.5}{\milli\second} of Doppler cooling using a combination of \SI{397}{\nano\meter}, \SI{866}{\nano\meter} and \SI{854}{\nano\meter} lasers. The latter is kept on \SI{30}{\micro\second} longer to empty the D$_{5/2}$ manifold (repumping). Doppler cooling is followed by a \SI{550}{\micro\second} monochromatic \SI{393}{\nano\meter} Raman laser pulse \textemdash used as a part of a sample-and-hold system to stabilize the laser power \textemdash, with an additional \SI{15}{\micro\second} AOD warm-up time, followed by another \SI{50}{\micro\second} of Doppler cooling and \SI{30}{\micro\second} \SI{854}{\nano\meter} repumping pulse. The ions are then optically pumped into the state $\ket{S} = \ket{4^2S_{1/2, m_j=-1/2}}$ with a \SI{30}{\micro\second} pulse of left circularly polarized  \SI{397}{\nano\meter}, light propagating along the quantization axis set by the applied magnetic field, combined with simultaneously starting a \SI{60}{\micro\second} repumping pulse at \SI{866}{\nano\meter} to empty the D$_{3/2}$ manifold. \\

\noindent\textbf{Photon generation (BCMRT).}
Ten attempts are made to detect a photon at the remote detection station that is entangled with a single trapped ion. Each attempt consists of 

\begin{itemize}
	\item \SI{30}{\micro\second} of repumping the D$_{5/2}$ and D$_{3/2}$ with 854 and \SI{866}{\nano\meter} laser pulses.
	\item \SI{30}{\micro\second} of optical pumping at \SI{397}{\nano\meter}  followed by another \SI{30}{\micro\second}  of repumping with 854 and \SI{866}{\nano\meter} lasers.
	\item A \SI{50}{\micro\second} photon generation Raman pulse at \SI{393}{\nano\meter} with \SI{15}{\micro\second} AOD warm-up time (simply referred to as `Raman pulse' henceforth).
	\item A \SI{210}{\micro\second} wait time. This wait time is sufficient for the photon to travel to the remote detection station and for the corresponding optical signal to be sent back to the lab upon detection of a single photon. 
	If a photon is successfully detected in the target time window, then a "success" TTL is sent to the ion control system.
\end{itemize}

\noindent\textbf{Measurement.}
If a photon is detected and the success-indicating TTL is received, the sequence continues with \SI{2.5}{\milli\second} ion-state readout. For ion-state readout, \SI{397}{\nano\meter} and \SI{866}{\nano\meter} laser light is shone onto the ion while fluorescence photons scattering at \SI{397}{\nano\meter} are detected with a photomultiplier tube (PMT). \\

\subsection{Ion-photon entanglement infidelity model}\label{appendix:ion-photon infid}

Here we describe our model of the effect of background photon detector counts on the measured ion-photon states. 
The model is used to explain the infidelities of the states $\rho^{\text{B}}_{\text{ip}}$ and $\rho^{\text{B}}_{\text{ip,rep}}$ in the main text. 
The model uses three inputs.
The first input is the rate of the background counts, defined as  detector clicks that didn't result from a photon from the ion. The background count rate (BCR) is extracted from the ion-photon state tomography experiments by looking far outside the time windows in which the photons from the ions are expected to arrive and summing the contributions from both detectors. 
The second input is the $P_{\text{photon}}$: the probability of detecting a photon that contributed to the measured ion-photon state ($\rho^{\text{B}}_{\text{ip}}$ or $\rho^{\text{B}}_{\text{ip,rep}}$).
The third input is an ion-photon state $\rho^{\text{input}}$. 
The model predicts what Bell fidelity is expected if the state $\rho^{\text{input}}$ was measured in the presence of the background counts and with the photon detection probability $P_{\text{photon}}$. 

The modeling is performed as follows. 
First, from the BCR we calculate $P_{\text{bg}}$: the average probability of a background count occurring in the photon detection window. 
Second, the measurement outcome probabilities expected for the state $\rho^{\text{input}}$ during the full state tomography are calculated. 
Third, these probabilities are renormalized such that their average equals  $P^\prime_{\text{photon}} = P_{\text{photon}}-P_{\text{bg}}$: the background-subtracted probability to detect a photon that belongs to the measured state.
Fourth, the average probability of a background count $P_{\text{bg}}$  is added to the probabilities obtained at the previous step. 
Fifth, a new density matrix is reconstructed via maximum likelihood tomography from the result of the previous step. 
Finally, the Bell fidelities  are calculated from this new density matrix.

As described in the main text, for $\rho^{\text{B}}_{\text{ip}}$ we use the BCR value of 2.04 cps and 
our best estimate of the entangled state injected into the QFC setup as $\rho^{\text{input}}$. 
For the state $\rho^{\text{B}}_{\text{ip,rep}}$ we use the BCR value of 2.06 cps and the ideal Bell state $\rho^{\text{input}} = \ket{\psi(0)}\bra{\psi(0)}$ as the input.

\subsection{Simulated photon wavepacket}\label{appendix:simulated wavepacket}
\label{sims1}

We obtain the simulated temporal single-photon wavepacket, presented as the dashed red curve in Fig.~\ref{fig:ion-photon}(a), in three steps. First, we calculate probability densities for the emission of a photon into the cavity from a single ion for chosen time bins using the Python package described in Chapter 3 of \cite{Fioretto2020}. The package solves a master equation describing the ion-cavity interaction. It operates in a Hilbert space comprising 18 electronic states of the ion and two frequency-degenerate modes of the cavity. The initial state for the numerical integration is $\ket{4^2S_{1/2,-1/2}}$; see Chapter 3 of \cite{Brandstatter2013} for a similar example using an earlier version of the code. From this first step, $P_1$ is obtained by summing the probability densities over all time bins. Next, we multiply each probability density by the total detection-path efficiency $P_\text{path} = 3.1 \times 10^{-3}$ (Sec. \ref{ion-photon entanglement}; details of $P_2$ to $P_8$ in Appendix \ref{appendix:path_eff}). Finally, we apply the ratio $P_\text{ip}/P_\text{ip,calc} = 1.13 \times 10^{-3} /(1.8 \times 10^{-3}$) (see Table \ref{table:eff budget}). \\
We now give the experimental parameters that we use as inputs to the Python package. We use a cavity decay rate of $\kappa = 2\pi \times 126$\;kHz (Sec.~II G in \cite{Canteri2025}) and a cavity resonance frequency jitter of $2\pi \times 10$\;kHz as reported in Ref.~\cite{SchuppKrcmarskyKrutyanskiyMeranerNorthupLanyon2021}. The orientation of the Raman laser path, the cavity, and the magnetic field (4.132\;G) matches the description in Sec.~I B of Ref.~\cite{KrutyanskiyCanteriMeraner2023}. 
From AC Stark shift measurements, we determine that the Rabi frequency of the $\sigma^{-}$-polarized component, driving the relevant carrier transition from $\ket{4^2S_{1/2,-1/2}}$ to $\ket{4^2P_{3/2,-3/2}}$, is $\Omega^{-} = 2\pi \times 31.8$\;MHz and the Rabi frequency of the $\sigma^{+}$-polarized component is $\Omega^{+} = 2\pi \times 18.4$\;MHz. 
The splitting of $\Omega^{-}$ into the two components of the bichromatic drive is calculated as described in Sec.~I B of Ref.~\cite{KrutyanskiyCanteriMeraner2023}. 
As reported in Chapter 4 of \cite{SchuppPhDthesis2021}, the maximum single-ion cavity coupling strength in our system is $g = 2\pi \times 1.53$\;MHz. The coupling strength of the photon generation process for an ion is proportional to the product of $g$ with $\Omega^{-}$. As in Sec. II C in \cite{Canteri2025}, we introduce a reduction to this product to account for coupling between the Raman laser and the Doppler-cooled ion string's thermal motion. Here, the reduction is given by a factor 0.84, corresponding to a Lamb-Dicke factor of $\eta = 0.09$ and phonon number $n = 10$ for both radial modes. Thus the photon generation product is $g \Omega^{-} \times 0.84$. 
\\ 

\subsection{Remote photon path efficiency}\label{appendix:path_eff}
The model for photon generation presented in Appendix \ref{appendix:simulated wavepacket} predicts the
probability for single-photon emission into the cavity, $P_1 = 0.59$. A total detection path efficiency factor $P_{\mathrm{path}}$ captures all losses that the photons subsequently face, including the imperfect detectors. Here, a detailed efficiency budget (resulting in $P_{\mathrm{path}} = 3.1(4)\times 10^{-3}$) is presented for the photon detection path. The symbols for the efficiencies of individual parts are the same as in Table \ref{table:eff budget}, except where $P_3 = P_{3.1}P_{3.2}P_{3.3}$. $P_{\mathrm{path}}$ is given by the product of $P_2$ through $P_8$. 
\begin{itemize}
	\item $P_2 = 0.43(3)$: The measured probability for a photon to exit the cavity into free space through the cavity output mirror after being emitted by a \ch{^{40}Ca^+} ion into the cavity. See $P_{\mathrm{esc}}$ in section I.G. in the supplemental material of Ref.~\cite{Canteri2025}. 
	\item $P_{3.1} = 0.96(1)$: The transmission of free-space optical elements that are between the cavity output mirror and the first fiber coupler. This was measured when the setup was built \cite{SchuppKrcmarskyKrutyanskiyMeranerNorthupLanyon2021}.
	\item $P_{3.2} = 0.83(3)$: The efficiency of coupling the photons into the first single-mode fiber. It is measured as the ratio of the 854 nm laser power at the end of the coupled fiber and that in front of the collimator. 
	\item $P_{3.3} = 0.77(3)$: The measured transmission through a lossy fiber joiner that joins the 854 nm photon fiber to the QFC input fiber. Measured using an \SI{854}{\nano\meter} laser and a powermeter with an additional patch fiber instead of the QFC input fiber. 
	\item $P_4 = 0.22(1)$: QFC external efficiency. It is the measured probability that an \SI{854}{\nano\meter} photon entering the QFC setup (Fig.~\ref{fig:detailledQFCsetup}) exits fiber-coupled at \SI{1550}{\nano\meter} and is described in Appendix \ref{appendix:QFC Setups and Characterization}. It involved the measurement of the \SI{854}{\nano\meter} laser power after the \SI{854}{\nano\meter} fiber collimator at the input of the QFC setup, and the power of the converted and fiber-coupled \SI{1550}{\nano\meter} laser light at the output of the QFC setup.
	\item $P_5 = 0.192(3)$: Channel Q transmission efficiency at \SI{1550}{\nano\meter}, determined by measuring the input and output power of a \SI{1550}{\nano\meter} laser sent through Channel Q. 
	\item $P_6 = 0.54(2)$: FBG transmission efficiency. This value corresponds to the maximum transmission measured through the fiber Bragg grating (FBG). The measurement was performed using a powermeter after tuning the FBG to the converted \SI{1550}{\nano\meter} laser light generated by the QFC setup.
	\item $P_7 = 0.62(2)$: Transmission efficiency of the photon polarization analysis board in Location B. It was measured using \SI{1550}{\nano\meter} laser light injected to the board. The input power was recorded immediately before the input coupler that connects the patch fiber from the FBG to the analysis board (see Fig.~\ref{figA1_setup}). The output power was measured separately for the transmission and reflection channels at the ends of the fibers connected to the detectors. 
	\item $P_8 = 0.82(2)$: The mean efficiencies of the two remote detectors, 0.81(3) and 0.83(3), which were measured by the manufacturer before shipping to Location B.
\end{itemize}

\section{Repeater-node demonstration details}\label{appendix:repeater}

In this appendix section, further details are provided on our repeater-node experiment. The results of that experiment were presented in Fig.~\ref{figure5}  and first discussed in Sec. \ref{Repeater node demonstration}. 

\subsection{Pulse sequence}\label{appendix:protocol} 
\label{repeaterprotocol}

\noindent The repeater-node protocol consists offour steps: 1. Initialization; 2. Loop 1; 3. Loop 2; 4. Deterministic Bell state measurement (DBSM).
Each step is given in more detail below. Wait times on a microsecond timescale are neglected. \\

\noindent\textbf{Initialization.}
Initialization begins with \SI{4}{\milli\second} of Doppler cooling using a combination of \SI{397}{\nano\meter}, \SI{866}{\nano\meter} and \SI{854}{\nano\meter} lasers. The latter is used for emptying the D$_{5/2}$ manifold (repumping).
Doppler cooling is followed by a \SI{550}{\micro\second} monochromatic \SI{393}{\nano\meter} Raman laser pulse \textemdash used as a part of a sample-and-hold system to stabilize the laser power \textemdash followed by another \SI{30}{\micro\second} of Doppler cooling and \SI{854}{\nano\meter} repumping pulse. The ions are then optically pumped into the state $\ket{S} = \ket{4^2S_{1/2, m_j=-1/2}}$ with a \SI{25}{\micro\second} pulse of left-circularly polarized \SI{397}{\nano\meter} light propagating along the quantization axis set by the applied magnetic field, 
combined with a simultaneous \SI{866}{\nano\meter} repumping pulse. The outer radial center-of-mass (COM) mode is then cooled down close to the motional groundstate by \SI{4}{\milli\second} of sideband-resolved groundstate cooling on the \SI{729}{\nano\meter} quadrupole transition.
Groundstate cooling is followed by another round of optical pumping at \SI{397}{\nano\meter}.

Up to this point, the cavity is filled with \SI{806}{\nano\meter} laser light used for locking the cavity length. 
Now, that \SI{806}{\nano\meter} light is turned off and the piezo actuators used to control the cavity length are held at constant voltage via a sample-and-hold system.
After the initialization sequence, both ions are prepared in the electronic state $\ket{S}$, and the outer radial COM mode is cooled down close to the groundstate. 
The other five motional modes are at Doppler-cooled temperatures. \\

\noindent\textbf{Loop 1.} In Loop 1, 20 attempts are made to distribute ion-photon entanglement over the \SI{17}{\kilo\meter} metropolitan fiber link. 
The sequence that is looped consists of 

\begin{itemize}
	\item A \SI{30}{\micro\second} repumping pulse at \SI{854}{\nano\meter} and \SI{866}{\nano\meter} which empty the D$_{5/2}$ and D$_{3/2}$ manifolds, followed by \SI{25}{\micro\second} optical pumping at \SI{397}{\nano\meter} that starts simultaneously with \SI{55}{\micro\second} long \SI{866}{\nano\meter} and \SI{854}{\nano\meter} laser pulses, which gives an extra \SI{30}{\micro\second} of repumping.
	\item If there is \SI{806}{\nano\meter} light inside the cavity, it is turned off (sample-and-hold).
	\item A \SI{60}{\micro\second} Raman pulse at \SI{393}{\nano\meter} on ion 1 (plus \SI{15}{\micro\second} AOD warm-up).
	\item A \SI{210}{\micro\second} wait time. This wait time is sufficient for the photon to travel to the remote detection station and for the corresponding optical signal to be sent back to the lab upon detection of a single photon. 
	If a photon is successfully detected in the target time window, then a ``success" TTL pulse is sent to the ion control system.
\end{itemize}
If a photon is successfully detected, Loop 1 is aborted and Loop 2 begins. 
If there is no photon detected, the sequence is iterated up to 20 times, and after every 5th iteration, \SI{0.4}{\milli\second} of sideband cooling on the outer radial COM mode is done. In the case of 20 sequential failed iterations (photon distribution attempts) in Loop 1, the \SI{806}{\nano\meter} light inside the cavity is turned on again, and the cavity length is stabilized.\\

\noindent\textbf{Loop 2.} Loop 2 is triggered when a photon is successfully detected at the remote detection station during Loop 1. 
Loop 2 starts with two sequential $\pi$ pulses (\SI{4}{\micro\second} and \SI{9}{\micro\second} \SI{729}{\nano\meter} pulses) to map the metastable qubit $\ket{D}$/$\ket{D'}$ down into the groundstate qubit $\ket{S}$/$\ket{S'}$, which serves as the quantum memory. These pulses are global (act on both ions equally) and therefore prepare the ion that was not used for photon generation in Loop 1 in $\ket{D}$/$\ket{D'}$. The rest of Loop 2 consists of the following sequence of pulses, which is iterated up to 22 times. 

\begin{itemize}
	\item A \SI{30}{\micro\second} repumping pulse at \SI{854}{\nano\meter} and \SI{866}{\nano\meter} which empties the D$_{5/2}$ and D$_{3/2}$ manifolds. Optical pumping is not performed within Loop 2, as (using our current method) it would destroy the state that is stored in the memory by pumping $\ket{S'}$ into $\ket{S}$.
	\item If there is \SI{806}{\nano\meter} light inside the cavity, it is turned off (hold in sample-and-hold is engaged).
	\item A \SI{60}{\micro\second} Raman pulse on ion 2 (plus \SI{15}{\micro\second} AOD warm-up).
	\item A spin-echo on the groundstate qubit by a \SI{112}{\micro\second} $\pi$-pulse with an alternating sign of the phase (sequentially implementing Pauli X and -X operations). 
\end{itemize}

If the photon generation on ion 2 (Loop 2) is successful in an odd number of attempts, an additional spin-echo is performed such that the total number of spin-echoes is always even. After jumping to the even- or odd branch in the sequence, the \SI{806}{\nano\meter} light inside the cavity is turned on again, and the cavity length is stabilized.
The sequence then proceeds with the DBSM part. \\

\noindent\textbf{Deterministic Bell state measurement (DBSM).}
The DBSM requires an entangling gate to be applied to the ions; the employed gate further requires the state to be encoded in the optical qubit $\ket{S}$/$\ket{D}$.
Therefore we start the DBSM with a single-ion-addressed operation which we refer to as ``addressed recombination", where ion 1 is mapped back from $\ket{S}$/$\ket{S'}$ to $\ket{D}$/$\ket{D'}$ by again doing two sequential composite $\pi$-pulses of lengths \SI{4}{\micro\second} and \SI{9}{\micro\second}, while leaving ion 2 untouched.
Details on the addressed recombination (and composite pulses) can be found in Section I.E.1. of the supplemental material of Ref.~\cite{KrutyanskiyCanteriMeraner2023}.  The rest of the DBSM proceeds as follows: 
\begin{itemize}
	\item A \SI{4}{\micro\second} \SI{729}{\nano\meter} $\pi$ pulse on the transition $\ket{D'} \leftrightarrow \ket{S}$, leaving the qubits encoded in the $\ket{S}$/$\ket{D}$ states.
	\item A \SI{205}{\micro\second} \SI{729}{\nano\meter} laser-driven M\o{}lmer-S\o{}rensen (MS) gate on both ions. Details on the MS gate can be found in Appendix \ref{ms_gate}.
	\item The \SI{806}{\nano\meter} cavity-locking light is injected into the cavity (sample in sample-and-hold is engaged).
	\item \SI{2.5}{\milli\second} of ion-qubit readout via electron shelving. 
	Here, \SI{397}{\nano\meter} and \SI{866}{\nano\meter} laser light are shone onto the ion string while fluorescence photons scattering at \SI{397}{\nano\meter} are imaged onto an EMCCD camera. 
	The camera resolves the fluorescence state of the ions individually.
\end{itemize}
The repeater-node protocol is hereby complete.

\subsection{Entangling M\o{}lmer-S\o{}rensen gate}\label{ms_gate}

\noindent For the DBSM, the two ions are projected onto a basis of maximally entangled states, this is done by first applying an entangling M\o{}lmer-S\o{}rensen (MS) gate, followed by projectively measuring the individual ionqubits in the logical basis by electron shelving. The effective measurement is described by the set of four operators that project into an orthogonal set of maximally entangled states $\{U^\dagger\ket{v}\bra{v}U\}$, where $\ket{v}$ are the four logical basis states of two ion qubits and $U=\exp\left(-i\sigma_\phi^1\sigma_\phi^2\pi /4\right)$ is the ideal action of the MS gate, where $\sigma_\phi^k=\cos(\phi)\sigma_\text{x}^k+\sin(\phi)\sigma_\text{y}^k$. The operators $\sigma_{\text{x(y)}}^k$ correspond to the Pauli x(y) operator on the $k^{th}$ qubit, while $\phi$ is controlled by the relative phase of the laser fields driving the MS gate. For example, obtaining the outcome $\ket{00}$ corresponds to an effective projection of the state of the ions before the MS gate into the maximally entangled state given by $U^\dagger\ket{00}=\frac{1}{\sqrt{2}}\left(\ket{00}+i\ket{11}\right)$, for a phase of $\phi=0$.

The MS gate works by illuminating the two qubits encoded in the optical states $\ket{S}/\ket{D}$ with a \SI{205}{\micro\second} off-resonant bichromatic \SI{729}{\nano\meter} laser pulse. The two tones lie outside of the respective outer radial red and blue COM sideband modes by $\delta=\SI{5091}{\hertz}$. The sideband frequency corresponds to $f_{\text{sb}}=\SI{2.00113(3)}{\mega\hertz}$. The two tones generate a qubit-state dependent force on the ions by off-resonantly driving the outer radial COM-mode. The two tones also shift the optical transition frequency via the AC-Stark effect, mostly due to far off-resonant dipole transitions. This AC-Stark shift is compensated by a third \SI{729}{\nano\meter} laser beam which illuminates the ions through a separate path (axial 729 access), which is blue-detuned from $\ket{S}$ to $\ket{D}$ by \SI{200}{\kilo\hertz}. 

The MS gate is partially characterized by determining the quality (fidelity) with which it can prepare a maximally entangled Bell-like state of two ions. Specifically, we prepare two ions in the product state $\ket{00}=\ket{SS}$, apply the gate to prepare the state $\rho$, and calculate the fidelity with the ideal state $\ket{\xi(\phi)}=\frac{1}{\sqrt{2}}\left(\ket{00}-ie^{2i\phi}\ket{11}\right)$. The state fidelity $F_{MS}=tr\left(\ket{\xi(\phi)}\bra{\xi(\phi)}\rho\right)$ is determined by measuring the logical populations and the amplitude of the parity oscillations, as described in  \cite{MonzEtAl2011, Kirchmairthesis}. Measurement of $F_{MS}$ is one of the calibrations interleaved with execution of the repeater node protocol, the average value over the six obtained values is 0.963(10), where the uncertainty is given by the range.

\subsection{Characterization of the ion-photon states stored in memory}\label{appendix:memorycost}
We conducted two separate experiments (on the same day, a few hours before the repeater node demonstration) to characterize how storage of ion 1 qubit in memory influences the corresponding ion-photon state fidelity. In the first experiment, we attempted to generate photons from two co-trapped ions sequentially, with 20 attempts on each ion. In the second experiment, photon generation on ion 2 was conditioned on detecting a photon from ion 1; after a successful ion 1 emission, ion 1 was stored in the ground-state memory qubit under our dynamical decoupling sequence while attempting to generate cavity photons from ion 2 until success. The number of attempts was up to 20 for ion 1 and 22 for ion 2. This second experiment is identical to the repeater-node protocol, but without the MS-gate implementation. Photons from both ions were detected at Location A along the same optical path in both experiments. The measured average Bell fidelities of the state of ion 1 and photon are $F_1$ = 0.959(5) in the first experiment and $F_2$ = 0.899(7) in the second, implying an infidelity of $F_1-F_2$ = 0.060(9) caused due to memory storage (error source B in Table \ref{table:remote ionphoton infid budget}).

\subsection{Ion-photon state infidelity due to polarization drifts}\label{appendix:repeater_ip_infid_pol_drift}

Section~\ref{Repeater node demonstration} presents our estimate of the contribution of drifts in fiber birefringence to the ion-photon state infidelity (error source D, Table~\ref{table:remote ionphoton infid budget}).
Here we provide details on how that estimate was obtained.
As described in Section~\ref{Repeater node demonstration}, polarization stabilization runs were interleaved with data-taking intervals for the repeater node protocol.
The $i^{\mathrm{th}}$ polarization stabilization run yields the nearest unitary rotation $U_i$ implemented by the fiber on the polarization of transmitted light, at the end of the preceding data-taking interval.
The extent to which the fiber birefringence has changed at the end of an interval is captured by comparison of $U_i$ with $U_{i-1}$.
At the beginning of that interval, fiber birefringence compensation has just been performed, and we expect that an injected perfect ion-photon Bell state $\ket{\Phi^{+}}$ would be transmitted at that time with a Bell fidelity of 1.
By the end of the interval, we calculate the reduction in transmitted Bell fidelity as $F_i=\left|\langle\Phi_i|\Phi_{i-1}\rangle\right|^2$, where $\ket{\Phi_k}=U_k\ket{\Phi^+}$ and $U_k$ acts on the photonic subsystem.
Our estimate for the average Bell fidelity over that interval---ending in the $i^{\mathrm{th}}$ polarization stabilization run---is the average $\bar{F}_i=(1+F_i)/2$.
Finally, to obtain an average Bell fidelity over all $n$ data-taking intervals, we calculate $\bar{F}_{\mathrm{total}}=\sum_{i=1}^{n}\bar{F}_i/n$, exploiting that the interval times are all roughly constant.
The result is $\bar{F}_{\mathrm{total}}=0.986(6)$, corresponding to an infidelity of $1-\bar{F}_{\mathrm{total}}=0.014(6)$.\\

\subsection{Simulated wavepackets}\label{repeatersims}
 
The simulated temporal single-photon wavepackets shown in Fig.~\ref{figure5}(a) are obtained in three steps, as described in Appendix~\ref{appendix:simulated wavepacket}. There is, however, one difference in the simulation of the wavepacket for Location A: We do not do optical pumping within Loop 2, as described in Appendix \ref{appendix:protocol}. As a consequence the initial state of ion 2 is modeled as an equal statistical mixture of $\ket{4^2S_{1/2,-1/2}}$ and $\ket{4^2S_{1/2,+1/2}}$. The wavepacket for Location B is shown as the dashed red curve in Fig.~\ref{figure5}(a)(i), whereas that for Location A is shown as the dashed blue curve in Fig.~\ref{figure5}(a)(ii). \\
Hereafter, the subscripts A and B denote Locations A and B, respectively. The modeled probability densities per time bin are multiplied by the path efficiencies $P_\text{path,B} = 3.3 \times 10^{-3}$ and $P_\text{path,A} = 11.2 \times 10^{-2}$ respectively, followed by applying the ratio of the measured to independently-estimated values for $P_\text{B,1}$, $1.4 \times 10^{-3}/(2.0 \times 10^{-3})$, and $P_\text{A,2}$, $4.0 \times 10^{-2}/(5.8 \times 10^{-2})$. The path efficiency for Location B is computed as the product of $P_2$-$P_8$ in Table \ref{table:eff budget}, with the following modifications (see Sec. \ref{Repeater node demonstration}): An additional photon switch with transmission 0.64 is included, and the QFC efficiency is 0.36 instead of 0.22. For Location A, the path efficiency is the product of $P_2$ and $P_3$ in Table \ref{table:eff budget}, the photon-switch transmission of 0.64, the photon analysis board transmission at \SI{854}{\nano\meter} of 0.80, and the SNSPD efficiency (0.82; Appendix \ref{appendix:setup-location A}). The numbers for the measured values for $P_\text{B,1}$ and $P_\text{A,2}$ can be found in Sec. \ref{Repeater node demonstration}. The numbers for the independently-estimated values for $P_\text{B,1}$ are calculated by the product $P_\text{path,B} \times P_1^1$ and for $P_\text{A,2}$ by $P_\text{path,A} \times P_1^2$. Both $P_1^1$ and $P_1^2$ can be found in Sec. \ref{Repeater node demonstration}.\\
Most experimental parameters for the simulation are provided in Appendix~\ref{appendix:simulated wavepacket}. Below, only the parameters that differ are listed:

\begin{itemize}
    \item \textbf{Location B:}
    \begin{itemize}
        \item Rabi frequency of the $\sigma^{-}$-driven transition: $\Omega_\text{B}^{-} = 2\pi \times 34.0$ MHz
        \item Rabi frequency of the $\sigma^{+}$-driven transition: $\Omega_\text{B}^{+} = 2\pi \times 19.7$ MHz
    \end{itemize}

    \item \textbf{Location A:}
    \begin{itemize}
        \item Rabi frequency of the $\sigma^{-}$-driven transition: $\Omega_\text{A}^{-} = 2\pi \times 32.7$ MHz
        \item Rabi frequency of the $\sigma^{+}$-driven transition: $\Omega_\text{A}^{+} = 2\pi \times 18.9$ MHz
    \end{itemize}
\end{itemize}

\subsection{Robustness of spin-echo protected ion-memory to \SI{50}{\hertz} noise}\label{appendix:modelSpinecho}

Here we present a model of how magnetic field noise at \SI{50}{\hertz} affects the ground-state memory qubit during our dynamical decoupling sequence. 
The implemented $X{/}{-}X$ sequence during Loop 2 of the repeater protocol consists of up to $N_{\mathrm{tot}}=22$ $\pi$ pulses (spin echoes) with alternative axes of rotation ($X, -X, X, -X$... etc), each of duration $t_{\mathrm{pulse}}$ and equally spaced by a wait time $t_\mathrm{w}$. 
The  \SI{50}{\hertz} magnetic field is modeled to interact with the ground-state qubit via the linear Zeeman effect, causing a time-varying frequency detuning between the applied spin-echo radio frequency (RF) and the $\ket{S}\leftrightarrow\ket{S'}$ transition of the form 
\begin{equation}
	\Delta(t) = \frac{\mu_\text{B}gB_0}{2\hbar}\sin{(2\pi50 t + \beta)},
\end{equation}
where $\beta$ is a phase, $g = 2$ is the Land\'e factor for the $\Ss$ manifold and $\mu_\text{B}$ is the Bohr magneton.  In a single run of the model (a single shot), the phase $\beta$ is considered to be fixed.

The detuning  $\Delta(t)$ results in an imperfection in the $n^{\mathrm{th}}$ spin-echo $\pi$-pulse. The imperfect action of the $n^{\mathrm{th}}$ spin-echo is modeled via the unitary rotation 
\begin{equation}
R_n = e^{-i\frac{\pi}{2}\bm{\sigma}\cdot\bm{c_n}},
\end{equation}
where we have introduced the Pauli vector $\bm{\sigma} = [\sigma_\text{x}, \sigma_\text{y}, \sigma_\text{z}]$ and the rotation axis $\bm{c_n}$. 
The rotation axis can be expressed as $\bm{c_n} = [(-1)^{n{+}1}, 0, \Delta(t_n)/\Omega]$, where $\Delta(t_n)$ is the detuning at the time of the middle of the $n^{\mathrm{th}}$ spin-echo, the spin-echo pulse length $t_{\mathrm{pulse}}=\pi/\Omega$ and $\Omega$ is the Rabi frequency of the RF spin-echo drive. We are thus effectively digitizing the magnetic field fluctuations in the model, which is reasonable since the length of each spin-echo (and wait time) is short compared to the period of \SI{50}{\hertz} (\SI{100}{\micro\second} $<<$ \SI{20}{\milli\second}). 

During the wait time that occurs between the $n^{\mathrm{th}}$ and $(n+1)^{\mathrm{th}}$ spin-echo, the detuning causes a $\sigma_\text{z}$ rotation of the form
\begin{equation}
W_n(t_{\mathrm{w}}) = e^{-i\Delta(\frac{t_{n+1}-t_{n}}{2}) t_{\mathrm{w}}\sigma_\text{z}/2},
\end{equation}
where the relevant detuning is fixed to the value at the middle of the wait time. 

The total unitary evolution of up to $N_{\text{tot}}$ spin echos with interleaved wait times is given by 
\begin{equation} \label{keymetho:eq:unitaryxy16}
U = W_n(t_{\mathrm{w}})\prod_n^{N_{\mathrm{tot}}} R_nW_n(t_{\mathrm{w}}).
\end{equation}
For a given $N_{\text{tot}}$, fifty unitaries $U_m$ are found, each using a different phase value $\beta$ of the \SI{50}{\hertz} magnetic field fluctuation sampled from a random uniform distribution. 
The corresponding single qubit process matrix $\chi$ is determined via  
\begin{equation}
	\label{eq:chi_average}
	\chi_{N_{\mathrm{tot}}} = \frac{1}{50} \sum_{m=1}^{50} \chi(U_m).
\end{equation}

In the repeater node protocol, after ion 1 emits a photon that is subsequently detected, Loop 2 proceeds with repeated photon-generation attempts on ion 2 (see Appendix \ref{appendix:protocol}). After each attempt, a spin-echo pulse is applied to the ground-state memory qubit of ion 1. If a photon is successfully detected in Loop 2 after an odd number of attempts, an additional $\pi$ pulse is inserted to ensure that the required $X/{-}X$ spin-echo sequence is completed.
The maximum number of photon-generation attempts in Loop 2 is 22, resulting in up to eleven distinct echo realizations. The possible numbers of attempts are therefore grouped into the eleven pairs: (1,2), (3,4), ..., (21,22). The pair (1,2) corresponds to an effective sequence containing one $X/{-}X$ echo, the pair (3,4) corresponds to a sequence containing two $X/{-}X$ echoes, and so on. For each realization, the corresponding single-qubit process matrix is calculated using the model described above, as given in Eq.~(\ref{eq:chi_average}), and is denoted by $\chi_r$. The effective quantum process is then calculated as a weighted mixture of these process matrices:
\begin{equation}
	\mathcal{E}_{\chi}
	=
	\frac{\sum_{r=1}^{11} p_r \, \chi_r}{\sum_{r=1}^{11} p_r}
\end{equation}
Here, $p_r$ denotes the photon counts detected from ion 1 recorded for the corresponding pair of Loop 2 attempts and are given by $p_r$ = [2602, 2420, 2248, 1968, 1753, 1522, 1447, 1303, 1159, 1093, 1106]. The process $\mathcal{E}_{\chi}$ is applied to one of the qubits of an ideal Bell state $\rho_{\mathrm{Bell}} = \ket{\Phi^{+}}\bra{\Phi^{+}}$
to obtain the output state
\begin{equation}
	\rho_{\mathrm{out}} = (\mathcal{E}_{\chi}\otimes\mathbb{I})\rho_{\mathrm{Bell}}.
\end{equation}
The Bell-state fidelity is then computed as
\begin{equation}
	F = \langle \Phi^{+} | \rho_{\mathrm{out}} | \Phi^{+} \rangle.
\end{equation}
The infidelity $(1-F)$ quantifies the contribution of the \SI{50}{\hertz} magnetic-field noise, during the dynamical-decoupling sequence, to the ion-photon state error when the ion qubit is stored in memory. We used this model with the parameters of the repeater node demonstration, $t_{\mathrm{pulse}}$ = \SI{111.9} {\micro\second}, $t_{\mathrm{w}}$ = \SI{235.8}{\micro\second} and magnetic-field noise amplitude \SI{36}{\micro\gauss}, which yields an infidelity of 0.03 in the ion-photon state. For a noise amplitude of \SI{54}{\micro\gauss}, the model yields an infidelity of 0.06.

\subsection{Estimation of the time to establish stored entanglement between distant ions via a QRN.}\label{appendix:rates_distant}
In Sec. \ref{Conclusion_and_outlook}, we discuss a 1D prototype repeater-node chain with the central QRN remaining in the lab and two QRNs deployed in opposite directions, across and outside of a city.
In that section the average time to establish stored entanglement between the ions in the two deployed QRNs via the central QRN using the two-photon click protocol is calculated as the inverse of the expression (S4) from \cite{KrutyanskiyCanteriMeraner2023}.
The value of $L/c$ entering that expression is taken to be \SI{180}{\micro\second}, corresponding to the round-trip light-travel time between any QRN and one of its neighbouring detection stations. 
For the parameter $P_s$ from \cite{KrutyanskiyCanteriMeraner2023} we use the two-photon heralding probability, namely
$P_s = P^2_{\text{B},1}/2=1\cdot 0.5\times 10^{-4}$.
We obtain the average time value of  \SI{5.4}{\second}. 
The average coherence time is calculated using (S7) from \cite{KrutyanskiyCanteriMeraner2023} using \SI{180}{\micro\second} photon travel time as the duration of each attempt and the same value of $P_s$ as above. We obtain the value of  \SI{3.6}{\second}.

For the single-photon click scheme, the maximum fidelity $F$ to establish entanglement between the deployed nodes via the central QRN is estimated as $F = F^2_0$, where $F_0$ is the fidelity bound for the entangled state of two ions entangled via the single-photon click protocol.
We now use the single-click heralding probability as $P_s = (1-F_0)P_{\text{B},1}=(1-\sqrt{F})P_{\text{B},1}$. Substituting $F = 0.9$, $P_{\text{B},1} = 0.01$ we obtain a factor of 5 higher value of $P_s$ and entanglement distribution rates compared to the two-click scheme.

\bibliographystyle{apsrev4-2}

\bibliography{Master_Bib_File}

\label{LastBibItem}

\end{document}